\documentclass[aps,prb,
               reprint,
               a4paper,
               amsmath,amssymb,superscriptaddress,showpacs]{revtex4-1}

\usepackage[utf8]{inputenc}
\usepackage[english]{babel}
\usepackage{graphicx}
\usepackage{feynmp}
\usepackage{bm}

\renewcommand{\Im}{\mathop{\mathrm{Im}}\nolimits}
\providecommand*{\bedt}{$\alpha$-(BEDT-TTF)$_{2}$I$_{3}$}
\providecommand*{\braket}[3]{\left \langle #1 \vert #2 \vert #3 \right\rangle}
\providecommand*{\spa}{\sigma_{\parallel}}
\providecommand*{\spe}{\sigma_{\perp}}
\providecommand*{\ce}{\hbar\omega_{c}}
\providecommand*{\cer}{\hbar\omega_{c}^{*}}
\DeclareMathOperator{\sgn}{sgn}

\DeclareMathOperator{\Tr}{Tr}

\begin{document}

\title{Longitudinal conductivity of massless fermions with tilted Dirac cone
       in magnetic field}

\date{\today}

\author{Igor \surname{Proskurin}}
\email{iprosk@hosi.phys.s.u-tokyo.ac.jp}
\affiliation{Department of Physics, The University of Tokyo, 
             Bunkyo-ku, Tokyo 113-0033, Japan}
\affiliation{Institute of Natural Sciences, Ural Federal University,
             Ekaterinburg 620083, Russia}
             
\author{Masao \surname{Ogata}}
\affiliation{Department of Physics, The University of Tokyo, 
             Bunkyo-ku, Tokyo 113-0033, Japan}

\author{Yoshikazu \surname{Suzumura}}
\affiliation{Department of Physics, Nagoya University,
             Nagoya 464-8602, Japan}
             
\begin{abstract}
    We investigate a longitudinal conductivity of a two-dimensional relativistic 
    electron gas with a tilted Dirac cone in magnetic field.
    It is demonstrated that the conductivity behaves differently in
    the directions parallel and perpendicular to the tilting of the cone. 
    At high magnetic fields, the conductivity at non-zero Landau levels in 
    the direction perpendicular to the tilting modifies non-trivially, in 
    contrast to the parallel case.
    At zero temperature, the crossover of the conductivity at the Dirac point from high 
    to low magnetic field is studied numerically.
    It is found that the tilting produces anisotropy 
    of the conductivity which changes with the magnetic field
    which is different from the anisotropy coming from the Fermi velocity.
    We also discuss the conductivity at finite temperatures and finite magnetic fields
    which can be directly compared with the experiments in \bedt{} organic conductor.
    We find that the tilting does not affect so much the magnetic-field dependence
    of the conductivity except for the prefactor. 
    We discuss the interpretation of recent experimental data and make some proposals
    to detect the effect of the tilting in future experiments.
\end{abstract}


\maketitle

\section{Introduction}
\label{secI}

    During the last decade, Dirac fermions have gained much attention in condensed 
    matter physics being of a great interest from both fundamental and applied 
    points of view.\cite{Fukuyama2012}
    The most prominent material where existence of massless Dirac particles was 
    unambiguously shown is graphene.\cite{Novoselov2005}
    Later Dirac electrons were theoretically predicted and experimentally observed in 
    many other materials including surface states of two- and three-dimensional 
    topological insulators,\cite{Qi2011} \bedt{} organic conductor,
    \cite{Tajima2006,Katayama2006,Kino2006,Kobayashi2007,Kajita2014}
    and BaFe$_{2}$As$_{2}$ iron-pnictide.\cite{Richard2010}
    The presence of relativistic Dirac quasiparticles in these materials
    gives rise to such interesting physical phenomena as unconventional 
    quantum Hall effect,\cite{Novoselov2005} 
    crossover from the positive to negative interlayer 
    magnetoresistance,\cite{Osada2008,Morinari2010} and giant 
    Nernst--Ettingshausen effect.\cite{Lukyanchuk2011,Proskurin2013,Konoike2013}

    Organic conductor \bedt{} is a layered material which consists of conducting layers of 
    BEDT-TTF molecules separated by insulating layers of I$_{3}^{-}$ anions. 
    Under the hydrostatic pressure about 1.5~GPa, it undergoes a transition 
    to the zero-gap state with a typical ratio of the in-plane to the 
    interlayer conductivity $\sim 10^{3}$ which makes this compound 
    a quasi-two-dimensional conductor.\cite{Tajima2009} 
    Unusual transport properties\cite{Tajima2006} in \bedt{}  were  explained by 
    finding the existence of the Dirac fermions in the  analytical band structure 
    calculation,\cite{Katayama2006} and  numerical one.\cite{Kino2006}
    Kobayashi  \textit{et al.} \cite{Kobayashi2007} showed that Dirac quasiparticles 
    should be described by an anisotropic tilted Weyl equation.
    Further experimental and theoretical studies provided confirmation
    of the massless Dirac nature of the carriers in this compound.
    \cite{Osada2008,Tajima2009,Morinari2010,Tajima2010,Sugawara2010}
    In contrast with graphene, where the carrier concentration is easily tuned by applying
    the gate voltage, the carrier number in \bedt{} is always fixed near the charge 
    neutrality point which makes difficult direct observation of the Landau levels.
    Recently, multilayer quantum Hall effect and Subnikov--de~Haas oscillations were
    observed in this compound using the method of contact electrification from 
    the substrate.\cite{Tajima2013}

    A notable feature of the massless fermions in \bedt{} is considerable 
    tilting of the Dirac cone in the energy-momentum space which breaks both 
    the Lorentz invariance and the chiral symmetry.
    The properties of systems with a tilted Dirac spectrum and their realizations
    in real materials were studied by several authors.
    \cite{Goerbig2008,Goerbig2009,Kawarabayashi2011, Sari2014, Trescher2015}
    It was found that in magnetic field the effect of tilting on the energy spectrum  is
    similar to the effect of an electric field on relativistic Landau levels in graphene. 
    \cite{Lukose2007,Goerbig2009}
    In the latter case, using the argument based on the Lorentz covariance of 
    the Dirac Hamiltonian, it was shown that the electric field squeezes the
    separation between the Landau levels and shifts the positions of the 
    wave functions in the real space.\cite{Lukose2007}
    Kawarabayashi \textit{et al.} \cite{Kawarabayashi2011} found that even if 
    the tilting is introduced, there exists the generalized chiral
    symmetry which provides existence and stability of the zero-mode 
    which governs the transport properties near the charge
    neutrality point in high magnetic fields. \cite{Tajima2009,Sugawara2010}
        
    A general problem of the transport of massless Dirac fermions was studied
    previously in different contexts. 
    In relation to graphene, Shon and Ando \cite{Shon1998} 
    calculated the conductivity in the self-consistent Born approximation (SCBA), 
    and Peres \textit{et al.}\cite{Peres2006} considered the effect 
    of lattice defects and electron-electron interactions.
    Sharapov \textit{et al.}\cite{Sharapov2003} made a fully analytical
    calculation of the zero-temperature conductivity for gapped Dirac excitations
    in the context of $d$-wave superconductors, and the problem of Shubnikov--de~Haas oscillations
    was studied in detail by Gusynin and Sharapov.\cite{Gusynin2005}
    With respect to organic conductors, the effect of electron-electron 
    interactions on the in-plane conductivity was considered 
    by Morinari and Tohyama.\cite{Morinari2010a}
    The effect of tilting on the in-plane conductivity 
    in zero magnetic field was studied previously by the authors.
    \cite{Suzumura2014,Suzumura2014a}
    In weak magnetic fields the effect of tilting on the Hall conductivity
    has been studied previously by Kobayashi \textit{et al.}\cite{Kobayashi2008}

    The purpose of this paper is to answer several questions.
    The first question is how the tilting of the Dirac cone affects 
    the transport properties of two-dimensional Dirac fermions in magnetic field.
    The second question is how to provide an experimental confirmation of the
    tilting. According to the knowledge of the authors, at the present time, there is
    no experimental evidence of the Dirac cone tilting in \bedt{}.
    In this paper, we make a proposal how it can be confirmed 
    in future experiments by detecting the anisotropy in the conductivity.
    The third question is related to the scenario of the magnetotransport 
    in \bedt{} which is motivated by recent experiments.\cite{Monteverde2013,Ozawa2014}
    We note that at the present time the magneto-conductivity in \bedt{}
    is not completely understood. First-principles band-structure
    calculations show that, together with the Dirac cone, \bedt{}
    also has a heavy-hole band near the Fermi level \cite{Kino2006}
    which might participate in the conducting properties. Based on this argument,
    the magneto-conductivity of \bedt{} in Ref.~\onlinecite{Monteverde2013}
    was explained in the framework of quasi-classical two-carrier
    model where the first carriers are massless Dirac electrons,
    and the second are massive holes. However, taking into account recent experimental results
    \cite{Ozawa2014}, we propose an alternative explanation which is 
    based only on the Dirac type of carriers. We discuss a criterion 
    which can make a clear distinction between these two models.   
  
    The paper is organized as follows. 
    In Sec.~\ref{secII}, we develop the formalism.
    In Sec.~\ref{secIII}, we consider the effect of the Dirac cone tilting
    on the zero-temperature conductivity. In the limit of high magnetic fields, 
    we study how the tilting modifies the conductivity
    when the chemical potential is situated at the $n$th Landau level.
    It is demonstrated  that the conductivity in this case is determined by 
    the wave function of the $n$th Landau level.
    Next, we study how the conductivity at $\mu = 0$ depends on magnetic field and impurity
    concentration. From the previous studies \cite{Sharapov2003, Shon1998}
    we know that massless Dirac fermions at $\mu = 0$ possess universal conductivity
    which is magnetic field and impurity independent. We show that in the presence
    of the tilting this behavior, in general, changes, however, in
    one particular direction, the conductivity remains independent of
    impurities and magnetic field. We also explain that the tilting leads
    to the anisotropy in the conductivity which changes in magnetic field,
    which makes it different from another type of the anisotropy coming from
    the Fermi velocity.
    This fact can be useful for experimental confirmation of the tilting
    in the two-dimensional Dirac systems.
    Section~\ref{secIV} is devoted to the conductivity at finite temperatures.
    First, we demonstrate that a combination of a crossover
    from weak to high magnetic fields, and the Zeeman splitting
    of $n = 0$ Landau level result in characteristic two-step decrease
    of the conductivity in the magnetic field detected experimentally.\cite{Monteverde2013,Ozawa2014}
    Next, based on the 
    self-consistent Born approximation (SCBA) arguments, we show that the
    magnetic-field dependence of the Landau level broadening produces minimum
    in the magneto-conductivity which may be crucial for understanding 
    recent experiments. \cite{Ozawa2014}
    We also find that the tilting does not affect so much the magnetic-field dependence
    of the conductivity except for the prefactor.
    Section~\ref{secV} is reserved for the summary.

\section{Formulation}
\label{secII}

    \subsection{Model Hamiltonian}
    
    The Hamiltonian for the two-dimensional relativistic electron gas
    inside the layer of BEDT-TTF molecules in magnetic field (for a given
    valley) can be written in the form of the generalized Weyl Hamiltonian
    \cite{Katayama2006,Goerbig2008}
    \begin{equation}
        H = v_{F} \left ( \eta \Pi_{y} \sigma_{0} + \Pi_{x} \sigma_{x}
          + \Pi_{y} \sigma_{y} \right ),
    \label{eq:hamilt}
    \end{equation}
    where $v_{F}$ is the Fermi velocity, $\Pi_{i}=-i\hbar\nabla_{i}+e A_{i}$ 
    denotes a canonical momentum, $\sigma_{i}$ is the Pauli matrix ($i=x,y$), 
    with $\sigma_{0}$ being the unitary matrix, $A_{i}$ is a magnetic vector 
    potential, and $-e$ is an electron charge.
    Here, we consider the case of a tilted Dirac cone with the isotropic Fermi velocity 
    due to the fact that the angular dependence of the Fermi velocity is small
    in \bedt{}. \cite{Kajita2014} 
    We comment on the effect of the anisotropy of the Fermi velocity at the
    end of Sec.~\ref{secIII}.
    Without loss of generality, we imply that the Dirac cone is tilted 
    in the $y$-direction with $0\leq\eta < 1$ being a degree of the tilting.
    
    The eigenvalue problem for the Hamiltonian in Eq.~(\ref{eq:hamilt}) 
    in the Landau gauge $\bm{A}=\left(0,Bx,0\right)$ can be solved by algebraic 
    methods which gives the following spectrum \cite{Morinari2009, Morinari2010}
    \begin{equation}
        E_{n} = \sgn(n) \sqrt{2 e B \hbar v_{F}^{2} \lambda^{3} |n|},
    \label{eq:spectrum}
    \end{equation}
    and eigenfunctions
    \begin{equation}
        \Psi_{kn}(x,y) = \frac{e^{iky}}{\sqrt{\ell L}} \Phi_{n} 
        \left ( \frac{ x - x_{n} + \ell^{2} k}{ \ell } \right ),
    \label{eq:eigenfunc}
    \end{equation}
    where
    \begin{equation}
        \Phi_{n}(x) = \frac{\chi^{(-)} h_{|n|}(x) 
        - i \sgn(n) \chi^{(+)} h_{|n|-1}(x)}{\sqrt{2(2-\delta_{n0})(1+\lambda)}},
    \label{eq:phi}
    \end{equation}
    $n=0,\pm 1,\pm 2,\dots$, $\lambda=\sqrt{1-\eta^{2}}$, 
    $\ell = \sqrt{\hbar/eB}$ is the magnetic length, 
    $k$ and $L$ are the wave number and the size of the system in 
    the $y$-direction, where periodic boundary conditions are implied, 
    $x_{n} = -\sgn(n)\eta \ell \sqrt{2 |n|/\lambda}$,
    $\chi^{(+)}=\left (1+\lambda,-i\eta\right )^{\mathrm{T}}$
    [$\chi^{(-)}=\left ( i\eta,1+\lambda \right )^{\mathrm{T}}$] denotes the eigenvector 
    of a generalized chiral operator \cite{Kawarabayashi2011}
    $\gamma = \left ( \sigma_{z} - i \eta \sigma_{x} \right )/\lambda$
    corresponding to $+1$ ($-1$) eigenvalue,
    \begin{equation}
        h_{|n|}(x) = \frac{\lambda^{1/4}}{2^{|n|/2}\pi^{1/4}\sqrt{|n|!}} \exp\left (
        -\frac{\lambda}{2} x^{2} \right ) H_{|n|} \left ( \sqrt{\lambda} x \right ),
    \end{equation}
    and $H_{n}(x)$ denotes the $n$th Hermite polynomial;
    $\sgn(x)$ is defined as $-1$ for $x < 0$, $0$ for $x = 0$, and $+1$ for $x >0$.
    Each Landau level is multiply degenerated with respect to $k$ with the
    degeneracy factor $V/(2\pi\ell^{2})$ where $V$ is a two-dimensional area
    of the system.

    In what follows, we consider the Hamiltonian in Eq.~(\ref{eq:hamilt}) for a
    given valley index and spin projection. 
    The effect of the Zeeman interaction between the electron spin
    and the magnetic field can be taken into account by including the term
    $-g\tau_{z}\mu_{B}B/2$ where $\tau_{z}$ is the Pauli matrix of the
    real spin, $\mu_{B}$ is the Bohr magneton, and $g$ denotes the $g$-factor.
    For the sake of brevity, we will omit the Zeeman term in most of the
    formulas following and restore it when it is necessary.

    \subsection{Longitudinal conductivity}
    
    At zero temperature, the longitudinal part of the conductivity tensor 
    for an electron gas in the presence of static disorder can be calculated 
    using the Kubo-Bastin-St\v{r}eda formula \cite{Bastin1971,Streda1975}
    \begin{equation}
        \sigma_{ii}(0,\mu)= \frac{e^{2}\hbar}{\pi} \left \langle
        \Tr \left [ v_{i} \Im G(\mu) v_{i} \Im G(\mu) \right ]
        \right \rangle,
    \label{eq:sigma0}
    \end{equation}
    where $\mu$ is a chemical potential, the velocity operator is defined as
    $v_{i} = (i/\hbar)\left [H, r_{i}\right ]$, 
    $\Im G \equiv (i/2)\left [ G^{(+)} - G^{(-)} \right ]$,
    and $G^{(\pm)}(\epsilon)$ denotes 
    $\left ( \epsilon - \mathcal{H} \pm i\delta \right )^{-1}$.
    Here, $\mathcal{H} = H + V_{\mathrm{imp}}$ is a total Hamiltonian
    which includes an impurity potential $V_{\mathrm{imp}}$ (specified later).
    The angle brackets stand for the average over the impurity positions.
    As was shown previously, in the case of short-ranged scatterers
    the current vertex corrections in Eq.~(\ref{eq:sigma0}) vanish,
    \cite{Bastin1971,Shon1998} and hereafter we imply that the effect 
    of the impurities is included into the Green function self-energy.
    The conductivity at the finite temperature can be restored from the
    expression
    \begin{equation}
        \sigma_{ii}(T,\mu) = -\int_{-\infty}^{\infty} d \epsilon
        \frac{\partial f(\epsilon)}{\partial \epsilon} \sigma(0,\epsilon)
    \label{eq:sigmaT}
    \end{equation}
    where $f(\epsilon) = \left \lbrace1 + 
    \exp \left [(\epsilon - \mu)/k_{B}T \right ]  \right \rbrace^{-1}$
    is the Fermi--Dirac distribution function with the temperature $T$.
    
    Calculating the trace in Eq.~(\ref{eq:sigma0}) with respect to the 
    Landau level eigenfunctions we obtain
    \begin{equation}
        \sigma_{ii}(0,\mu) = \frac{e^{2} \hbar v_{F}^{2}}{2 \pi^{2} \ell^{2}}
        \sum_{m,n = -\infty}^{\infty} 
        \left \vert \braket{n}{\bar{v}_{i}}{m} \right \vert^{2}
       \mathcal{A}_{n}(\mu) \mathcal{A}_{m}(\mu),
    \label{eq:sigma01}
    \end{equation}
    where $i = x,y$,
    \begin{equation}
        \bar{v}_{x} = \sigma_{x} \quad \mbox{and} \quad \bar{v}_{y} = \sigma_{y}+\eta,
    \end{equation}
    and 
    \begin{equation}
        \mathcal{A}_{n}(\mu) = \frac{\Gamma(\mu)}
        {\left (\mu - E_{n} - R(\mu)\right )^{2} +
        \Gamma^{2}(\mu)}.
    \end{equation}
    We imply that the electron self-energy 
    $\Sigma(\mu)\equiv R(\mu)+i\Gamma(\mu)$ 
    is free of the Landau level index dependence.
    Here, $\sigma_{ii}$ denotes the conductivity per valley and per 
    one spin projection. 
    The Zeeman interaction in Eq.~(\ref{eq:sigma01}) can be easily 
    restored by the substitution 
    $\sigma_{ii}(0,\mu)\to\sigma_{ii}(0,\mu + \frac{g}{2}\mu_{B}B)+
    \sigma_{ii}(0,\mu - \frac{g}{2}\mu_{B}B)$.
    
    The matrix elements of the velocity operators can be calculated explicitly
    \begin{eqnarray}
        \braket{m}{\bar{v}_{x}}{n} &=& -i \lambda \Delta_{n,-m}
        P^{n}_{m}(\eta \Delta_{n,m}), 
    \label{eq:vx}
    \\
        \braket{m}{\bar{v}_{y}}{n} &=& \lambda^{2} \Delta_{n,m}
        P^{-n}_{m}(\eta \Delta_{n,m}),
    \label{eq:vy}
    \end{eqnarray}
    where $n \ne m$, $\Delta_{n,m} = \sgn(n)\sqrt{|n|} - \sgn(m)\sqrt{|m|}$,
    \begin{widetext}
    \begin{multline}
        P_{m}^{n}(\eta \Delta_{n,m}) = \frac{1}{\sqrt{(2-\delta_{n0})(2-\delta_{m0})}}
        \sqrt{\frac{|m|!}{|n|!}}
        \exp\left ( -\frac{1}{2} \eta^{2} \Delta_{n,m}^{2} \right )
        \left ( \eta \Delta_{n,m} \right )^{|n|-|m|-1} 						\\
        \times
        \left (L_{|m|}^{|n|-|m|}(\eta^{2} \Delta_{n,m}^{2}) - \sgn(n)\sgn(m)\sqrt{\frac{|n|}{|m|}}
        L_{|m|-1}^{|n|-|m|}(\eta^{2} \Delta_{n,m}^{2})\right ),
    \label{eq:Pnm}    
    \end{multline}
    \end{widetext}
    and $L_{m}^{\alpha}(x)$ denotes the generalized Laguerre polynomial. 
    The matrix elements of velocity operators with $n = m$ equal to zero.
    The details of the derivation are given in Appendix~\ref{Appendix_A}.
    
    For simplicity of the subsequent analysis, we introduce
    dimensionless conductivities in the direction perpendicular
    $\spe = \sigma_{xx}/\left (\frac{e^{2}}{\pi h}\right )$
    and parallel $\spa = \sigma_{yy}/\left (\frac{e^{2}}{\pi h}\right )$
    to the tilting where $h = 2\pi \hbar$. 
    The dimensionless conductivities can be written as functions
    of dimensionless parameters
    \begin{eqnarray}
        \spe(x) &=& \frac{1}{2\lambda} \sum_{n,m}
        \Delta_{n,-m}^{2} \left [ P_{m}^{n}(\eta \Delta_{n,m})\right ]^{2}
        g_{m}(x) g_{n}(x), 
    \label{eq:s_per}
    \\
        \spa(x) &=& \frac{\lambda}{2} \sum_{n,m} \Delta_{n,m}^{2}
        \left [P_{m}^{-n}(\eta \Delta_{n,m})\right ]^{2} g_{m}(x) g_{n}(x),
    \label{eq:s_par}
    \end{eqnarray}
    where
    \begin{equation}
        g_{n}(x) = \frac{\gamma}{(x-\sgn(n)\sqrt{|n|})^{2}+\gamma^{2}},
    \end{equation}
    $x = (\mu-R(\mu))/(\cer)$, $\gamma = \Gamma(\mu)/(\cer)$, and
    $\cer \equiv \lambda^{3/2} \ce$ is a cyclotron energy
    $\ce = \sqrt{2e\hbar v_{F}^{2} B}$ renormalized
    by the tilting.
    
    In the $\eta = 0$ limit the velocity operators in Eqs.~(\ref{eq:vx}) and (\ref{eq:vy})
    have matrix elements only for Landau levels with $|n| = |m| \pm 1$,
    and the expressions for the conductivities in Eqs.~(\ref{eq:s_per}) and
    (\ref{eq:s_par}) are considerably simplified.
    The conductivity becomes isotropic $\spa = \spe = \sigma_{0}$ where
    \begin{multline}
        \sigma_{0}(x) = \frac{\ce}{4} 
        \sum_{\alpha \alpha' = \pm 1}\sum_{n=0}^{\infty} 
        \frac{\gamma}{(x-\alpha \sqrt{n+1})^{2}+\gamma^{2}}\\ \times
        \frac{\gamma}{(x-\alpha'\sqrt{n  })^{2}+\gamma^{2}}.
    \label{eq:sigma02}
    \end{multline}
    The summation over the Landau levels in this case can be done
    analytically, \cite{Sharapov2003} which gives
    \begin{multline}
        \sigma_{0}(x)=
        1-\frac{x^{2}+\gamma^{2}}{1+16x^{2}\gamma^{2}} 
        \left[
        \frac{x^{2}+4x^{2}\gamma^{2}+8x^{2}\gamma^{2}(x^{2}
        +\gamma^{2})}{(x^{2}+\gamma^{2})^{2}} \right. \\ \left.
        -4 x\gamma \Im \psi\left( \left(\gamma + i x \right)^{2} \right)
       \vphantom{\frac{x^{2}}{v^{2}}} \right].
    \label{eq:sigma03}
    \end{multline}
    The details are given in Appendix~\ref{Appendix_B}.
    In the case when $R(\mu)=0$ and $\Gamma(\mu)$ is a constant (constant damping approximation), 
    the conductivity at low magnetic field can be approximated using the expansion
    $\psi(z) = \log z - 1/(2z) - 1/(12z^{2})+ \dots$
    which yields
    \begin{equation}
        \sigma_{xx}(0,\mu) \approx
        \frac{e^{2}}{\pi h} \left[ \Phi_{0}\left( \frac{\mu}{\Gamma} \right)-
        \frac{(\ce)^{4}}{32\Gamma^{4}} \Phi_{2}\left( \frac{\mu}{\Gamma} \right)
        \right] \; ,
    \label{eq:mf_expan}
    \end{equation}
    where $\Phi_{0}(x)$ and $\Phi_{2}(x)$ are defined by 
    \begin{equation}
        \Phi_{0}(x) = \frac{1}{2} \left[ 1 + \left( x + \frac{1}{x} \right)
        \tan^{-1} x \right] \; ,
    \end{equation}
    and
    \begin{multline}
        \Phi_{2}(x) = -\frac{1}{x^{2}} \left[ 
        \frac{1-x^{2}}{1+x^{2}} - \frac{1+x^{2}}{x} \tan^{-1}x  \right.
        \\ \left.
        + \frac{8 x^{2} (1-x^{2})}{3(1+x^{2})^{3}} \right].
    \end{multline}
    The expression for $\Phi_{0}$ is well known.\cite{Shon1998,Sharapov2003}
    In the $|x| \gg 1$ limit, we obtain $\Phi_{0}\approx \pi |x|/2$, $\Phi_{2} \approx \pi/(2|x|)$,
    while $\Phi_{0} = 1+x^{2}/3$, $\Phi_{2}=128 x^{2}/15$ for $x \ll 1$.

    \subsection{Self-consistent Born approximation}
    
\begin{figure}
    \includegraphics[scale=1.0]{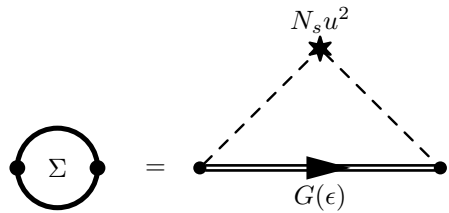}
    \caption{Diagrammatic representation of Eq.~(\ref{eq:scba}).
             Each dashed line on the right-hand side stands for 
             $u \delta(\bm{r}-\bm{R}_{j})$ 
             and the star means the average over $\bm{R}_{j}$ 
             and summation over $j$.}
    \label{fig:scba}
\end{figure}

    In order to take into account the effect of scattering on the 
    impurities in a self-consistent manner, we consider the 
    total Hamiltonian in the form of
    $\mathcal{H} = H + V_{\mathrm{imp}}$ where
    \begin{equation}
        V_{\mathrm{imp}} = u \sum_{j} \delta\left ( 
        \bm{r} - \bm{R}_{j} \right )
    \end{equation}
    describes the interaction between the conduction electrons and 
    randomly distributed point-like impurities at the positions
    $\bm{R}_{j}$ with the scattering potential $u$.
    Here, we imply that the impurity potential does not mix the 
    valleys, which corresponds to the case of short-ranged scatterers
    in Ref.~\onlinecite{Shon1998}.
    The self-energy $\Sigma(\epsilon)$ in SCBA can be calculated using the method 
    developed by Bastin \textit{et al.} \cite{Bastin1971} and Shon and Ando \cite{Shon1998}
    which leads to the self-consistent equation 
    $\Sigma(\epsilon) = N_{s}u^{2}V^{-1} \Tr G(\epsilon)$ illustrated in Fig.~\ref{fig:scba}.
    The explicit form for this equation reads
    \begin{equation}
        \Sigma(\epsilon) = \frac{N_{s}u^{2}}{2\pi\ell^{2}} \sum_{n=-\infty}^{\infty}
    \frac{1}{\epsilon - E_{n} -\Sigma(\epsilon)}.
    \label{eq:scba}
    \end{equation}
    where $N_{s}$ is the impurity concentration.
    Here, we ignore the effect of the Zeeman interaction for simplicity which is discussed below.
    
    In zero magnetic field, we can replace the summation over the Landau levels
    in Eq.~(\ref{eq:scba}) by integration. 
    In the Boltzmann limit, where we neglect the real part of $\Sigma(\epsilon)$
    and suppose that $\Gamma(\epsilon)=\Im \Sigma(\epsilon)$ is negligible, we obtain
    \cite{Shon1998}
    \begin{equation}
        \Gamma(\epsilon) = \frac{\pi \varkappa}{\lambda^{3}}|\epsilon|,
    \label{eq:boltzmann}
    \end{equation}
    where $\varkappa = N_{s}u^{2}/(2\pi v_{F}^{2}\hbar^{2})$ is a dimensionless parameter.
    The only difference of Eq.~(\ref{eq:boltzmann}) from the case of $\eta = 0$
    is the appearance of the factor $\lambda^{-3}$ which comes from the 
    angular dependence of the density of states at the Fermi level
    \citep{Goerbig2008,Suzumura2014}
    \begin{equation}
        \int_{0}^{2\pi} \frac{d \phi}{2\pi} \frac{1}{(1-\eta \sin \phi)^{2}}=
        \frac{1}{(1-\eta^{2})^{3/2}}=\frac{1}{\lambda^{3}}.
    \end{equation}
    In the opposite limit, when $\epsilon \to 0$ first, we obtain
    $\Gamma(0) = E_{c} \exp\left [-\lambda^{3}/(2\varkappa)\right ]$
    where we introduced the cut-off energy $E_{c}$ to regularize the 
    logarithmic divergence. 
    The only difference of this result from $\eta = 0$ case studied in
    Ref.~\onlinecite{Shon1998} is the factor $\lambda^{3}$.

    In the quantizing magnetic field, when $\epsilon \approx E_{N}$, 
    we can keep in Eq.~(\ref{eq:scba}) only the contribution with 
    $n=N$. In this case, $\Gamma(\epsilon)$ is approximated as \cite{Shon1998}
    \begin{equation}
        \Gamma(\epsilon) = \frac{\ce \sqrt{\varkappa}}{\sqrt{2}}
        \sqrt{1 - \frac{2}{\varkappa} \frac{(\epsilon-E_{N})^{2}}{ (\ce)^{2}}}.
    \label{eq:gammaN}
    \end{equation}
    
    In finite magnetic fields, in order to have a convenient analytic 
    expression for Eq.~(\ref{eq:scba}), we rewrite it in the following form:
    \begin{equation}
        \Sigma(\epsilon) = N_{s} u^{2} \int_{-\infty}^{\infty} d \omega
        \frac{D_{0}(\omega)}{\epsilon - \omega -\Sigma(\epsilon)}
    \label{eq:self_en}
    \end{equation}
    where
    \begin{equation}
        D_{0}(\omega) = \frac{1}{4 \pi \ell_{B}^{2}}
        \sum_{n=-\infty}^{\infty} 
        \delta\left (\omega - E_{n} \right )
    \end{equation}
    is the density of states in the absence of the impurities. 
    Equation~(\ref{eq:self_en})
    contains a divergence which can be treated analytically using the convenient 
    representation for $D_{0}(\omega)$ obtained in Ref.~\onlinecite{Sharapov2004}:
    \begin{multline} 
        D_{0}(\omega) = \frac{1}{4 \pi \ell_{B}^{2}}  \sgn \omega \frac{d}{d \omega}
        \left \lbrace \theta(\omega^{2}) \left [ \frac{\omega^{2}}{(\cer)^{2}}
        \right. \right. \\
        \left. \left.
        +\frac{1}{\pi} \tan^{-1}\cot \frac{\pi \omega^{2}}{(\cer)^{2}}
        \right ] \right \rbrace.
    \end{multline}
    Now the integration in Eq.~(\ref{eq:self_en}) can be performed explicitly 
    which finally gives the self-consistent equation
    \begin{equation}
        \frac{\Sigma}{\cer} = -\frac{\varkappa}{\lambda^{3}}
        \Phi \left ( \frac{\mu - \Sigma}{\cer}; \frac{E_{c}}{\cer} \right )
    \label{eq:scba0}
    \end{equation}
    where
    \begin{multline}
        \Phi(x;\epsilon_{c}) = x \left [ \log\left ( \epsilon_{c}^{2} - x^{2} \right ) -
        \psi(x^{2}) \right. \\
        \left.
        -\frac{1}{2 x^{2}} - \pi \cot (\pi x^{2}) \right ].
    \end{multline}
    The effect of the Zeeman splitting can be introduced in Eq.~(\ref{eq:scba0}) by changing 
    the right-hand side to
    \begin{equation}
        \sum_{\sigma = \pm 1} 
        \Phi \left ( \frac{\mu - \Sigma- \frac{g}{2}\sigma \mu_{B} B}{\cer}; \frac{E_{c}}{\cer} \right ).
    \label{eq:zeeman}    
    \end{equation}
    
    \begin{figure}
        \centerline{\includegraphics[scale=.75]{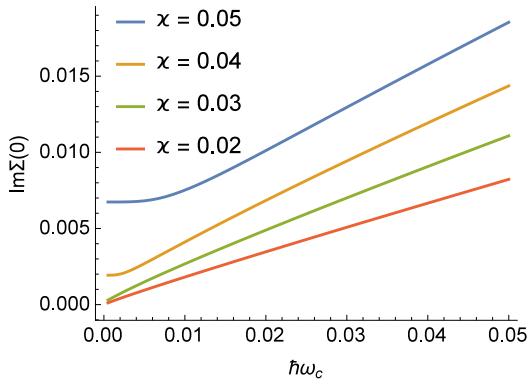}}
        \caption{(Color online) $\Im \Sigma(\epsilon)$ at $\epsilon = 0$ as a 
        function of $\ce$ for several $\varkappa$ with $E_{c} = 1$ and $\eta = 0$.}
    \label{fig:gamma}
    \end{figure}

    The $\epsilon$ dependence of $\Im\Sigma(\epsilon)$ was studied in Ref.~\onlinecite{Shon1998}.
    In Fig.~\ref{fig:gamma}, we show how $\Im\Sigma(\epsilon)$ depends on the 
    cyclotron energy at $\epsilon = 0$. 
    In order to obtain Fig.~\ref{fig:gamma}, we solved Eq.~(\ref{eq:scba0}) 
    numerically with fixed $E_{c} = 1$. 
    At $\ce = 0$ the value of $\Im\Sigma(0)$ for small $\varkappa$ becomes
    exponentially small in agreement with the expression 
    $E_{c} \exp\left [-\lambda^{3}/(2\varkappa)\right ]$.
    For large enough cyclotron energy, the dependence becomes linear
    which corresponds to $\Gamma(0) \sim B^{1/2}$. 
    In the following, we will denote 
    \begin{equation}
    \label{GammaB}
        \Gamma(B)=\Im \Sigma(0) \sim \sqrt{B}
    \end{equation}
    obtained in SCBA in order to distinguish it from the constant $\Gamma$.

\section{Conductivity at zero temperature}
\label{secIII}

    In this section, we analyze the effect of the tilting on 
    the zero temperature conductivity in high magnetic fields where the
    Landau quantization of the energy spectrum plays the principal role.
    At fist, we analyze the chemical potential dependence of the conductivity using
    Eqs.~(\ref{eq:s_per}) and (\ref{eq:s_par}). After that, we give a clear physical
    explanation of the conductivity at the $N$th Landau level and discuss 
    its relation for the quasi-classical picture. Second, we numerically
    study how the conductivity at the Dirac point depends on magnetic 
    field and impurity concentration. The results are given for two
    models of the impurity scattering: constant broadening approximation and SCBA.

    \subsection{High magnetic fields}
    \label{SecIIIA}
    
    We start with analyzing the conductivity at $T = 0$ in the high magnetic field
    when Landau levels are well separated ($\ce \gg \Gamma(\mu)$).
    In this case, when the chemical potential is close to $E_{N}$ we can keep 
    only the terms with $n = N$ in the summation in Eqs.~(\ref{eq:s_per}) 
    and (\ref{eq:s_par}) which gives
    \begin{equation}
        \sigma_{l}(\mu) = \sigma_{N}^{(l)} \frac{\Gamma^{2}_{N}(\mu)}
        {\left (\mu - E_{N}\right )^{2}+ \Gamma^{2}_{N}(\mu)},
    \end{equation}
    where $l = (\perp,\parallel)$.
    The coefficients $\sigma_{N}^{(l)}$, which are $\sigma_{l}(\mu)$ at $\mu = E_{N}$,
    are defined by
    \begin{eqnarray} \label{QNperp}
        \sigma_{N}^{(\perp)} &=& \frac{e^{2}}{\pi h\lambda}
        \sum_{m \ne N} \frac{\Delta_{N,-m}^{2}}{\Delta_{N,m}^{2}} 
        \left [ P_{N}^{m}(\eta \Delta_{N,m})\right ]^{2}, 
        \\ \label{QNpar}
        \sigma_{N}^{(\parallel)} &=& \frac{e^{2}\lambda}{\pi h}
        \sum_{m \ne N} \left [ P_{N}^{-m}(\eta \Delta_{N,m})\right ]^{2}.
    \end{eqnarray}
    For the constant broadening approximation the conductivity in the vicinity of the $N$th 
    Landau level is given by
    \begin{equation}
        \sigma_{l}(\mu) = \sigma_{N}^{(l)} \frac{\Gamma^{2}}
        {\left (\mu - E_{N}\right )^{2}+ \Gamma^{2}},
        \label{eq:sigmaN1}
    \end{equation}
    while if we use Eq.~(\ref{eq:gammaN}) for $\Gamma_{N}$ in SCBA, the expression for the 
    conductivity becomes
    \begin{equation}
        \sigma_{l}(\mu) = \sigma_{N}^{(l)} \left [
        1 - \frac{2}{\varkappa} \frac{\left (\mu - E_{N}\right )^{2}}{(\ce)^{2}} \right ].
    \label{eq:sigmaN}
    \end{equation}
    Note that $\sigma_{N}^{(l)}$ does not depend on $\Gamma(\mu)$, and the difference between
    constant $\Gamma$ and SCBA appears only in the $\mu$-dependence of $\sigma_{l}(\mu)$.
    
    The $\eta = 0$ case was studied by Shon and Ando \cite{Shon1998} who showed 
    that $\sigma_{N}^{(l)} = \frac{e^{2}}{\pi h}(\delta_{N0}+2|N|)$. 
    For finite $\eta$, using the numerical
    summation over Landau levels in Eqs.~(\ref{QNperp}) and (\ref{QNpar}),  
    we have found
    similar results
    \begin{eqnarray}
        \sigma_{N}^{(\perp)} &=& \frac{e^{2}}{\pi h\lambda}
        \left ( \delta_{N0} + 2|N| -N \eta^{2} \right ),
        \label{eq:QNper}
        \\
        \sigma_{N}^{(\parallel)} & = & \frac{e^{2}\lambda}{\pi h} \left ( \delta_{N0} + 2|N| \right ).
        \label{eq:QNpar}
    \end{eqnarray}
    Note that for $N = 0$, these equations can be verified explicitly using Eq.~(\ref{eq:Pnm}),
    \begin{equation}
        \lambda \sigma_{N=0}^{(\perp)} = \lambda^{-1} \sigma_{N=0}^{(\parallel)} \sim
        \sum_{n = 1}^{\infty} 
        \frac{\left ( \eta^{2} n \right )^{n - 1}}{n!} e^{-\eta^{2} n} = 1.
    \end{equation}
    while for $|N| > 1$ the expansions (\ref{app:vx_expand}) and (\ref{app:vy_expand})
    allow to confirm these equations up the order of $\eta^{2}$.

\begin{figure}
        \centerline{\includegraphics[scale=.75]{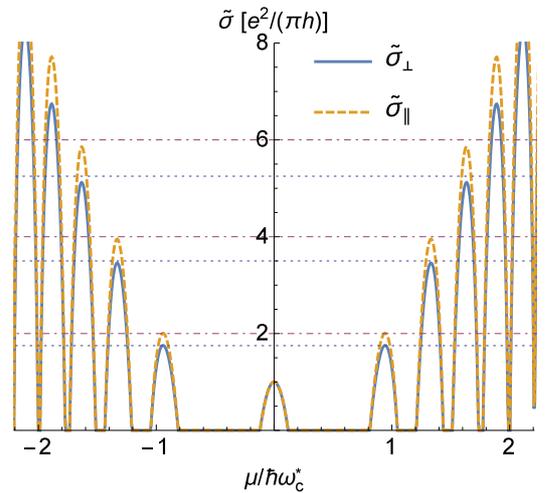}}
        \caption{(Color online) Chemical potential dependence of the normalized 
        conductivities $\tilde{\sigma}_{\parallel} = \lambda^{-1} \spa$ and 
        $\tilde{\sigma}_{\perp} = \lambda \spe$ in SCBA with $\eta = 0.5$,
        $\varkappa = 0.005$, and $E_{c}/\cer = 100$.
        The dotted-dashed (dotted) horizontal line is a guide for eyes at the value
        $2N$ ($2N-N\eta^{2}$).   }
    \label{fig:mu_dependence}
    \end{figure}

    From Eqs.~(\ref{eq:QNper}) and (\ref{eq:QNpar}) we find out that both
    $\spe$ and $\spa$ are modified by the tilting
    through the factors $\lambda^{-1}$ and $\lambda$, respectively, which
    can be accounted for the geometrical modification of the 
    quasi-classical electron orbit (discussed below). 
    However, $\sigma_{\perp}$ has an additional modification by the tilting 
    for $N \ne 0$ as demonstrated in
    Fig.~\ref{fig:mu_dependence}, where the normalized conductivities 
    $\tilde{\sigma}_{\parallel} = \lambda^{-1} \sigma_{\parallel}$ and
    $\tilde{\sigma}_{\perp} = \lambda \sigma_{\perp}$ are shown as functions
    of $\mu$ calculated numerically in SCBA using the summation over 50
    Landau levels in Eqs.~(\ref{eq:s_per}) and (\ref{eq:s_par}).
    
    The reason why $\tilde\sigma_{\parallel}$ at $\mu = E_{N}$ remains the same
    as in the $\eta=0$ case, while
    $\tilde\sigma_{\perp}$ is reduced by $N \eta^{2}/\lambda$ can be understood as follows. 
    From the Eq.~(\ref{eq:sigma01}), the conductivity at $\mu = E_{N}$ can be expressed as 
    \begin{equation}
    \sigma_{N}^{(l)} = \frac{e^{2} \hbar}{\pi^{2} \ell^{2}}
    \sum_{m}{}^{'} 
    \frac{\braket{N}{v_{l}}{m}\braket{m}{v_{l}}{N}}{\left (E_{N}-E_{m}\right )^{2}}
    \end{equation}
    where the prime indicates that the $m = N$ term is omitted. Using the relation
    $v_{l} = (i/\hbar)[H,r_{l}]$, we can write the off-diagonal matrix elements
    of the velocity operator as
    \begin{equation}
      \braket{N}{v_{l}}{m} = \frac{i}{\hbar} \left ( E_{N} - E_{m} \right )
      \braket{N}{r_{l}}{m} 
    \end{equation}
    which gives
    \begin{equation} \label{sigmaN}
    \sigma_{N}^{(l)} = 
    \frac{e^{2}}{ \pi^{2} \hbar }
    \frac{\langle \left (\Delta r_{l}\right )^{2} \rangle_{N}}{ \ell^{2}}
    \end{equation}
    where $\langle \left (\Delta r_{l}\right )^{2} \rangle_{N} \equiv
    \braket{N}{r_{l}^{2}}{N}-\braket{N}{r_{l}}{N}^{2}$. Even in the presence of
    the tilting, the zero-temperature conductivity  
    at $\mu = E_{N}$ in the quantum limit is determined only by 
    the $N$th Landau level wave function and proportional to the quantum-mechanical average of 
    $(\Delta r_{l})^{2}$ in the $\Psi_{kN}$-state.

\begin{figure}
        \centerline{\includegraphics[scale=.375]{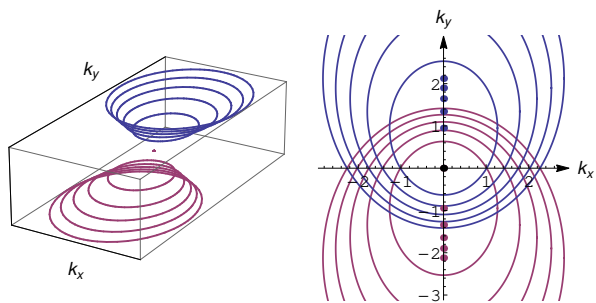}}
        \caption{(Color online) Semiclassical trajectories of a massless 
                 Dirac electron with the tilted cone in the $k$ space with $\eta = 0.6$ 
                 for $n=-5,\ldots,5$. The units of $k_{x}$ and $k_{y}$ are $\ell^{-1}$. 
                 The blue (purple) ellipses correspond to the upper (lower) band.
                 The dots show centers of the ellipses.
                 Each ellipse has its focus at the origin and its center
                 displaced by $x_{n}\ell^{-2}$ in the $y$-direction. 
                 The eccentricity of each ellipse is $\eta$.}
    \label{fig:trajectories}
    \end{figure}

    From the semiclassical point of view, the trajectories of a Dirac 
    electron in the momentum space with the tiling $0 < \eta < 1 $ 
    are displaced ellipses which can be parametrized as
    \begin{eqnarray}
    k_{x} &=& b_{N}\sin \phi,\\
    k_{y} &=&  x_{N}\ell^{-2} + a_{N} \cos \phi,
    \end{eqnarray}  
    with the semimajor (semiminor) axis equal to
    $a_{N} = \ell^{-1}\sqrt{2|N|/\lambda}$ ($b_{N} = \ell^{-1}\sqrt{2|N| \lambda}$)
    and $0 \le \phi < 2\pi$. 
    The area of each ellipse is independent of $\eta$ and given by 
    $S_{N}=2\pi |N|\ell^{-2}$. 
    Note that the Berry phase contribution to the semiclassical quantization 
    rule $\varphi_{B} = \pi$, as in the case without tilting. \citep{Goerbig2008} 
    Each ellipse has its focus at the axes origin and its center displaced
    by $x_{N}\ell^{-2}$ in the $k_{y}$-direction, as shown in Fig.~\ref{fig:trajectories}. 
    The eccentricity of each ellipse is $\eta$. 
    The trajectory in the real space can be obtained by rotation of the orbit in 
    the momentum space by $\pi/2$ and rescaling it with the factor $\ell^{2}$.
    If we make a semiclassical calculation of 
    $\langle \left (\Delta r_{l}\right )^{2} \rangle_{N}$, we obtain
    \begin{equation}
      \langle(\Delta x)^{2}\rangle_{N} = \int_{0}^{2\pi} \frac{d\phi}{2\pi} 
      \left (x - x_{N}\right )^{2}=
      \frac{|N|}{\lambda},
    \end{equation}
    and $\langle(\Delta y)^{2}\rangle_{N} = \lambda |N|$. The appearance of the factor
    $\lambda^{-1}$ ($\lambda$) in the $\sigma_{\perp}$ ($\sigma_{\parallel}$) is therefore
    accounted for the ellipticity of the electron orbit. However, in the semiclassical
    picture the displacement of the orbit has no effect on the conductivity and 
    behavior of $\sigma_{\perp}$ and $\sigma_{\parallel}$ 
    at $\mu = E_{N}$ is the same as in the case of $\eta = 0$.
    
    The quantum-mechanical average of the $r_{l}^{2}$ and $r_{l}$ ($l = x,y$) operators
    calculated with respect to the eigenfunctions in Eq.~(\ref{eq:eigenfunc}) gives the
    following results:
    \begin{eqnarray}
     \label{eq:x}
    &&\braket{N}{x}{N} = \frac{3}{2}x_{N},\\
    \label{eq:x2}
    &&\braket{N}{x^{2}}{N} = 
    2 x_{N}^{2} + \frac{\ell^{2}}{2\lambda}\left ( \delta_{N0} + 2|N| \right ),\\ 
    \label{eq:y2}
    &&\braket{N}{y^{2}}{N} = 
    \frac{\lambda\ell^{2}}{2}\left ( \delta_{N0} + 2|N| \right ),
    \end{eqnarray}
    and $\braket{N}{y}{N} = 0$ which justifies Eqs.~(\ref{eq:QNper}) and
    (\ref{eq:QNpar}) obtained by numerical calculations. These results are
    different from those obtained in the semiclassical picture, because
    $\braket{N}{x}{N} \ne x_{N}$. Note that according to our gauge choice,
    $y$ is an unbounded operator and a regularization procedure is required.
    The details are given in Appendix~\ref{Appendix_C}.

    \subsection{Magnetic-field dependence}
    
    It is well known that in the $\eta = 0$ case, the zero-temperature conductivity
    at $\mu = 0$ is independent of magnetic field and impurity 
    concentration and given by the universal value $e^{2}/(\pi h)$. \cite{Shon1998}
    The natural question is how this behavior is modified for $\eta > 0$.
    At zero magnetic field, we already know the answer.
    In our previous work, we have found that the conductivities
    at $\mu = 0$ are independent of the impurity broadening and defined by the following
    expressions: \cite{Suzumura2014}
    \begin{eqnarray}
        \spe & = & \frac{1}{\sqrt{1-\eta^{2}}}, \\
        \spa & = & \frac{\sin^{-1}\eta}{\eta}.
    \end{eqnarray}
    If we compare these  results with the case of strong magnetic field given by 
    Eqs.~(\ref{eq:sigmaN1}) and (\ref{eq:sigmaN}) with $N = 0$, we would find that $\spe$ remains
    the same in both limiting cases $\cer \to \infty$ and $\cer \to 0$, while $\spa$ increases
    from $\sqrt{1-\eta^{2}} < 1$ to $\sin^{-1} \eta/\eta > 1$ as $\cer$ is reduced
    from $\infty$ to zero.
    
\begin{figure}
        \centering
        \begin{minipage}{.45\textwidth}
            \centerline{\includegraphics[scale=.75]{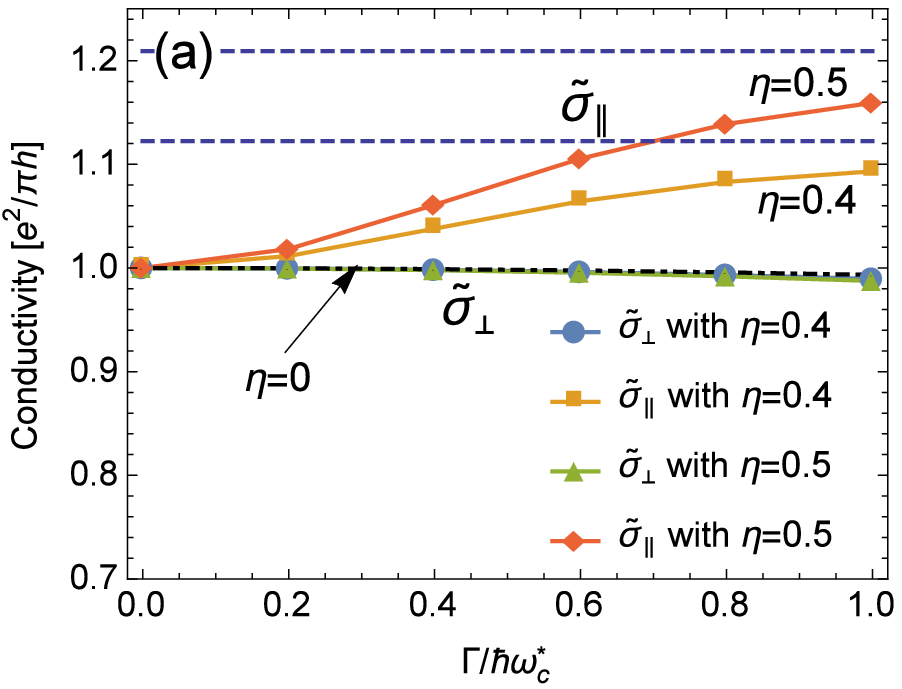}}
        \end{minipage}
        \begin{minipage}{.45\textwidth}
            \centerline{\includegraphics[scale=.75]{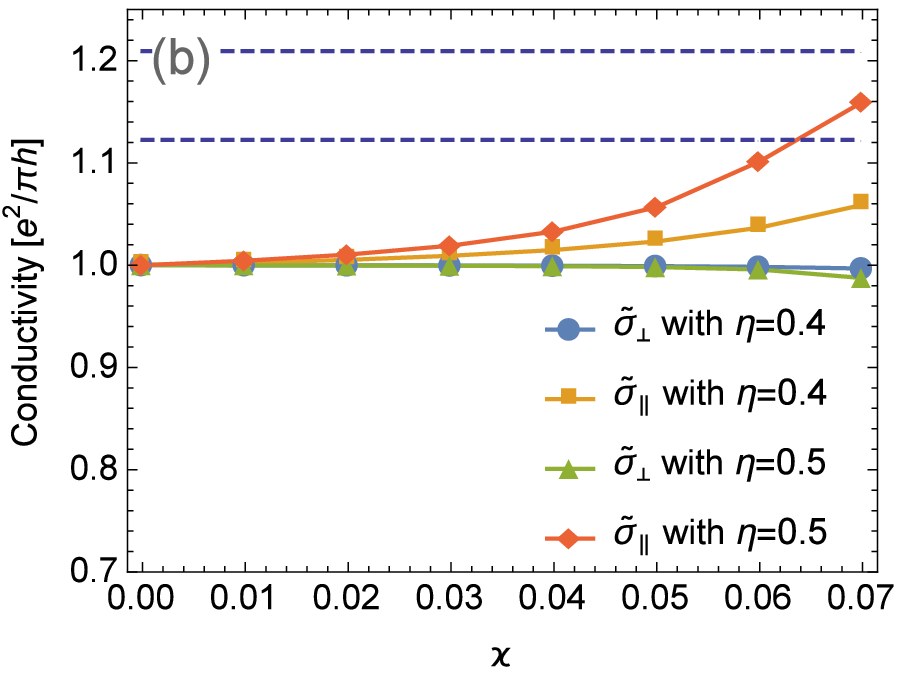}}
        \end{minipage}
        \caption{(Color online) Normalized conductivities $\tilde{\sigma}_{\parallel} = 
                     \lambda^{-1} \spa$ and $\tilde{\sigma}_{\perp} = \lambda \spe$ 
                     at $\mu = 0$ with $\eta = 0.4$ and $0.5$ calculated from 
                     Eqs.~(\ref{eq:s_per}) and (\ref{eq:s_par}) numerically 
                     using $N=150$ Landau levels. 
                     (a) The conductivities in the constant broadening
                     approximation as functions of $\Gamma/\cer$;
                     $\eta = 0$ case is indicated by the dotted-dashed line.
                     (b) The conductivities in SCBA as functions of
                     $\varkappa$ with $E_{c}/\cer = 100$.
                     In each figure, the upper (lower) dashed line corresponds to the limiting value 
                     $\sin^{-1}\eta/(\lambda \eta)$ at zero field
                     for $\eta=0.5$ ($\eta=0.4$).\cite{Suzumura2014} }     
    \label{fig:crossover}     
    \end{figure}

    First, we analyze the behavior of the conductivity at low magnetic fields in the 
    constant damping approximation.
    In order to study the behavior of the conductivities in the region of moderate 
    magnetic fields, we have made a numerical summation in Eqs.~(\ref{eq:s_per})
    and (\ref{eq:s_par}) over $N = 150$ Landau levels. 
    Figure~\ref{fig:crossover}~(a)
    shows normalized conductivities $\tilde{\sigma}_{\perp} = \lambda \spe$ and
    $\tilde{\sigma}_{\parallel} = \lambda^{-1} \spa$ at $\mu = 0$ for moderate 
    tilting $\eta = 0.4$ and $0.5$.
    Note that at $\mu = 0$ the only parameter is $\Gamma/\cer$ which is
    proportional to $B^{-1/2}$.
    Within the numerical accuracy we found that $\tilde{\sigma}_{\perp}$ 
    remains constant with increasing $\Gamma/\cer$ and does not depend on 
    $\eta$, while $\tilde{\sigma}_{\parallel}$
    increases with increasing $\Gamma$ or $\eta$, but remains smaller than the 
    corresponding limiting value $\sin^{-1} \eta/(\lambda \eta)$ shown
    by dashed lines. 
    Note that for large values of $\eta$, numerical calculations become complicated
    especially due to the inter-band matrix elements of velocity 
    operators with large $|m|$ and $|n|-|m|$ [see Eq.~(\ref{eq:Pnm})].

    Second, we show the normalized conductivities  
    in SCBA in Fig.~\ref{fig:crossover}(b) where Eq.~(\ref{eq:scba0}) has been used.
    For the self-consistent calculation we kept $E_{c}/\cer$ fixed and varied
    parameter $\varkappa$ which is proportional to the total number of 
    impurities. 
    In SCBA, the behavior of the conductivities in the crossover region remains 
    qualitatively the same as discussed above. 
    As $\varkappa$ increases, within the numerical accuracy  $\tilde{\sigma}_{\perp}$
    remains constant, while  $\tilde{\sigma}_{\parallel}$ increases, but remains
    below the limiting value at zero field (which is the same in both
    constant broadening and SCBA cases).
    
    At the end of this section, we comment about the 
    difference between the anisotropy of the conductivity given by the tilting
    and the anisotropy due to the difference of Fermi velocities 
    $v_{x} \ne v_{y}$ in the $x$ and $y$-directions. In the latter case, the energy spectrum
    is \cite{Goerbig2008} $E_{n} = \sgn(n)v_{F}^{*} \sqrt{2eB\hbar|n|}$, where
    $v_{F}^{*}=\sqrt{v_{x}v_{y}}$. The quasi-classical orbits are ellipses 
    centred at the axes origin with  the eccentricity defined by $\sqrt{v_{x}/v_{y}}$.
    Equation~(\ref{sigmaN}), in this case, gives the same conductivities at $\mu = E_{N}$ 
    as in the isotropic case renormalized by the factor $v_{x}/v_{y}$ ($v_{y}/v_{x}$) 
    for $\sigma_{xx}$ ($\sigma_{yy}$),
    which coincides with the conductivities obtained by the quasi-classical argument.
    Note that in the case of the anisotropy induced by $v_{x} \ne v_{y}$, the 
    ratio $\sigma_{xx}/\sigma_{yy} = v_{x}^{2}/v_{y}^{2}$ is independent of 
    the magnetic field (zero-field case was considered in Ref.~\onlinecite{Suzumura2014}),
    while if the anisotropy is produced by the tilting, $\sigma_{xx}/\sigma_{yy}$
    changes from $\sin^{-1}\eta/(\eta\lambda)$ in zero field to $\lambda^{-2}$ in
    the high-field limit. The magnetic-field-dependent anisotropy of the 
    in-plane conductivity which saturates in the quantum limit can be used
    as an evidence of the tilting in Dirac electron systems
    with $\mu$ close to the Dirac point.

\section{Conductivity at finite temperatures}
\label{secIV}

    In the following, we discuss the magnetic-field dependence at finite 
    temperature and the temperature dependence of the conductivity and 
    the resistivity at $\mu = 0$ which can be directly compared with experiments.
    \cite{Tajima2006,Monteverde2013,Ozawa2014}
    The experimental measurements of the magnetic-field dependence of the in-plane
    conductivity in \bedt{} organic conductor demonstrated a two-step decrease 
    of the conductivity with increasing the magnetic field.
    \cite{Monteverde2013,Ozawa2014}
    In Ref.~\onlinecite{Monteverde2013} this two-step decrease was interpreted 
    in terms of quasi-classical two-carrier model, where Dirac carriers and massive carriers coexist.
    Similar two-step behavior of the conductivity was also reported in Ref.~\onlinecite{Ozawa2014}.
    However, in Ref.~\onlinecite{Ozawa2014} detailed analysis of the 
    magnetic-field dependence with fixed $T$ revealed the existence
    of a novel minimum in the conductivity at a moderate magnetic field 
    which scales as $T^{2}$ in the temperature
    range of $1.5~\mathrm{K} < T < 5~\mathrm{K}$. In this section,
    taking account of the recent experiment,\cite{Ozawa2014} we show
    an alternative explanation for the two-step behavior
    in terms of one carrier scenario of Dirac electrons.
    Furthermore, in the present explanation, the minimum of
    conductivity can be understood. 
    At first, we consider the constant damping approximation
    which is frequently used in the literature. \cite{Kobayashi2008,Lukyanchuk2011,
    Sharapov2003,Sharapov2004, Gusynin2005}
    Secondly, we show how the results obtained in the constant damping approximation are 
    modified if SCBA for the impurity scattering is used.
    We start with analyzing $\eta = 0$ case in detail, and, after that,
    the effect of the tilting is considered.
    
    Hereafter, parameters are chosen with respect
    to the application to \bedt{}:\cite{Kajita2014} 
    $v_{F} = 5 \times 10^{4}$~m/s, $\Gamma/k_{B} = 3$~K, and $g=2$, 
    if these values are not indicated explicitly.
    
    \subsection{Conductivity in constant damping approximation}
    \label{SecIVA}
    
    \begin{figure}
    \centering
        \begin{minipage}{.45\textwidth}
            \centerline{\includegraphics[scale=.75]{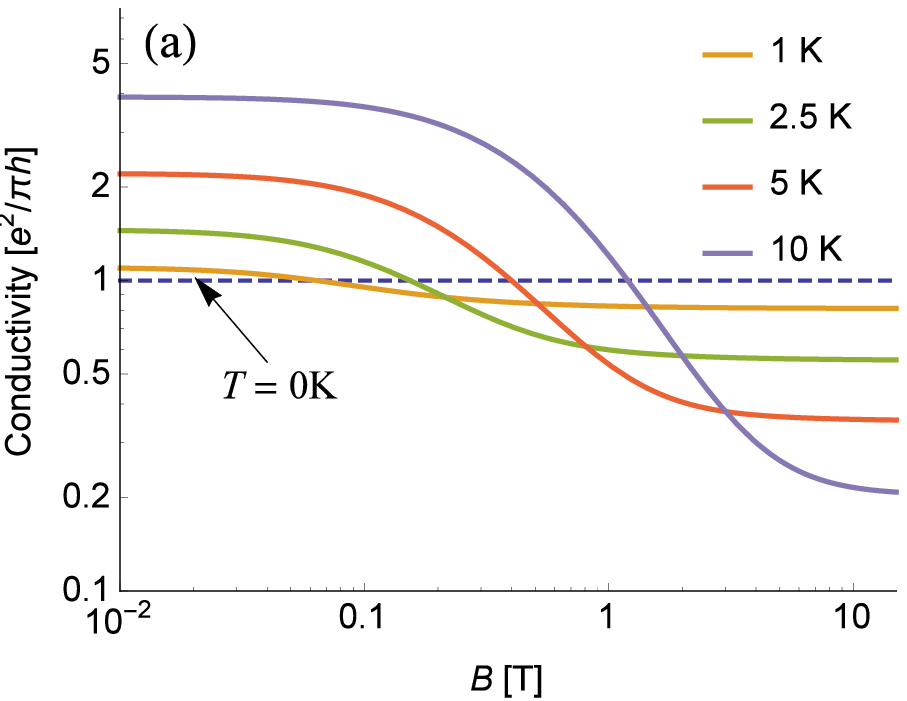}}
        \end{minipage}
        \begin{minipage}{.45\textwidth}
            \centerline{\includegraphics[scale=.75]{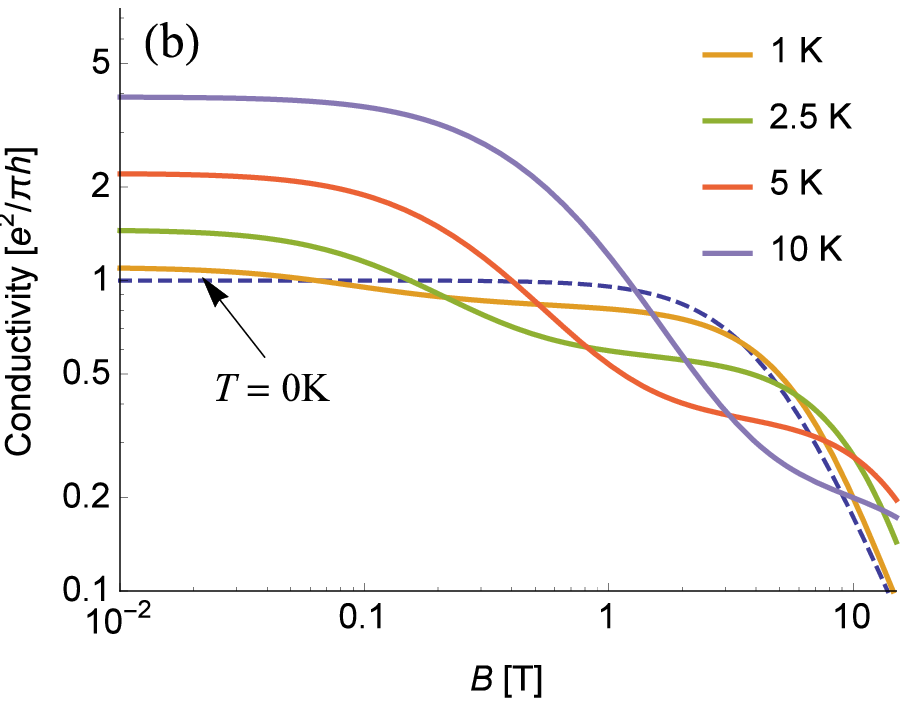}}
        \end{minipage}
        \caption{(Color online) Magnetic-field dependence of the conductivity 
                 at $\mu = 0$ in the constant broadening approximation for
                 several temperatures without the Zeeman interaction (a) and
                 with the Zeeman interaction included (b).}    
    \label{fig3}
    \end{figure}

    In this section, we consider a simple case where the self-energy has the form
    $\Sigma(\epsilon) = i \Gamma$, with $\Gamma$ being a phenomenological parameter. 
    Magnetic-field dependence of the conductivity calculated from 
    Eqs.~(\ref{eq:sigmaT}) and (\ref{eq:sigma03}) for $\mu = 0$ and $\eta = 0$
    is shown in Fig.~\ref{fig3}. 
    Figure \ref{fig3}(a) shows the result without the Zeeman interaction 
    and \ref{fig3}(b) with the Zeeman interaction.
    First, we discuss the case without the Zeeman interaction [Fig.~\ref{fig3}(a)].
    We interpret that the decrease of $\sigma$ is associated with the crossover 
    from the low magnetic-field
    region where Landau levels overlap to the quantum region
    where Landau levels are well separated.
    Actually, this crossover occurs when $\ce \sim k_{B}T$.
    In the limit of $B \to \infty$, the conductivity saturates at the value 
    determined by the $n = 0$ Landau level.

    Next, we discuss the temperature dependence.
    The effect of the temperature on the conductivity is the opposite
    in low and high magnetic fields.    
    In the low-field region, increasing of the temperature enhances the 
    conductivity. 
    This will be due to the thermal activation of the carriers.
    In contrast, in the quantum region, increasing the temperature reduces
    the conductivity. This is because the $n = 0$ Landau level has a 
    temperature smearing and, as a result, reduces the conductivity. 
        
    At low temperatures, the crossover in Fig.~\ref{fig3}~(a) between the 
    low magnetic-field region and high magnetic-field region can be discussed
    analytically from Eqs.~(\ref{eq:sigmaT}) and (\ref{eq:sigma03}). 
    We can use the Sommerfeld expansion for the Fermi--Dirac 
    distribution $-\partial f/\partial \epsilon = \delta(\epsilon-\mu)+(\pi^{2} k_{B}^{2}T^{2}/6)
    \delta''(\epsilon-\mu)$ which gives 
    \begin{equation}
        \sigma(T) = \frac{e^{2}}{\pi h} \left ( 1 + \frac{\pi^{2}k_{B}^{2}T^{2}}{6 \Gamma^{2}}  
        \sigma''(0)\right ),
    \label{eq:crossover}
    \end{equation}
    where $\sigma''(0)=d^{2}\sigma(0)/d(\mu/\Gamma)^{2}$ defined at $\mu = 0$.
    $\sigma''(0)$ can be found analytically from
    Eq.~(\ref{eq:sigma03})
    \begin{equation}
        \sigma''(0) = 
        -2 \left [ 1 + 4 \gamma^{2} + 8 \gamma^{4} - 8 \gamma^{6} \psi'(\gamma^{2}) \right ].
        \label{d2sigma}
    \end{equation}
    In large magnetic field ($\gamma \ll 1$), $\sigma''(0) \to -2$, 
    and the conductivity in Eq.~(\ref{eq:crossover}) 
    approaches $\frac{e^{2}}{\pi h}(1-\pi^{2}k_{B}^{2}T^{2}/3\Gamma^{2})$.
    This explains the increase of $\sigma(T)$ as a function of $T$ in high magnetic fields.
    When the magnetic field decreases, $\sigma(T)$ reaches the universal value $e^{2}/(\pi h)$
    at $\ce/\Gamma \approx 1.30$ which gives zero for the right-hand side in Eq.~(\ref{d2sigma}),
    and roughly represents the crossover between low-field and high-field regions.
    At low magnetic fields ($\gamma \gg 1$), $\sigma''(0) \approx \frac{2}{3} - \frac{8}{15}
    \gamma^{-4}$, 
    and the behavior of the conductivity is given by
    \begin{equation}
      \label{eq54}
    \sigma(T) \approx \frac{e^{2}}{\pi h}\left [ 1 + \frac{\pi^{2} k_{B}^{2} T^{2}}{9 \Gamma^{2}}
    \left ( 1 - \frac{4}{5} \frac{(\ce)^{4}}{\Gamma^{4}} \right ) \right ].
    \end{equation}
    This explains the increase of $\sigma(T)$ as a function of $T$ in low
    magnetic field.
    
    Next, we discuss the effect of the Zeeman interaction
    in Fig.~\ref{fig3}~(b). 
    When we include the Zeeman term there appears the second step decrease 
    in the high magnetic field due to the opening of the gap at $\mu = 0$.
    The magnetic-field dependence of the conductivity
    in this region can be understood as follows. At zero temperature, the conductivity per 
    spin near $\mu = 0$ in the spin-splitting region can be approximated as 
    \begin{equation}
        \sigma(0,\mu) = \frac{e^{2}}{2 \pi h} \sum_{s=\pm 1}
         \frac{\Gamma^{2}}{(\mu +gs\mu_{B} B/2)^{2} +\Gamma^{2}}.
     \label{eq:split}    
    \end{equation}
   At $\mu = 0$
    \begin{equation}
        \sigma(0,0) = \frac{e^{2}}{\pi h }
        \frac{1}{1 +  (g \mu_{B} B/2 \Gamma  )^{2}}.
    \label{eq:rho00}    
    \end{equation}
    This expression indicates that $\sigma(0,0)$ decreases when 
    $g \mu_{B} B > 2 \Gamma$. 
    At finite temperatures, $\sigma(T)$ is obtained by Eq.~(\ref{eq:sigmaT}).
    In this case, $\sigma(T)$ starts to decrease for sufficiently large magnetic field
    $2\mu_{B}B > k_{B}T$.
    
    \begin{figure}
        \centerline{\includegraphics[scale=.8]{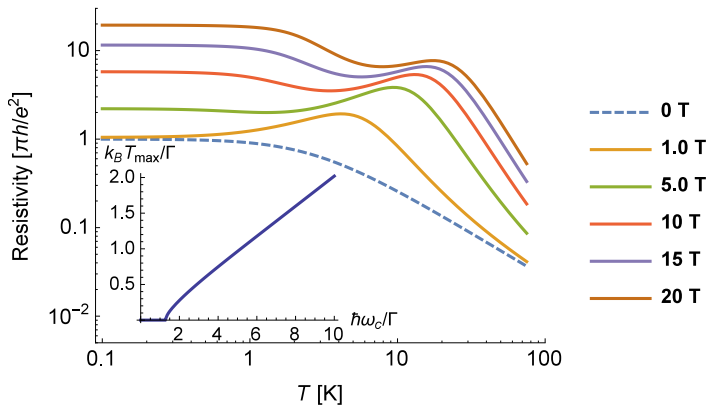}}
        \caption{(Color online) Temperature dependence of the resistivity at $\mu = 0$ 
                  for several magnetic fields.
                  The dashed line shows the resistivity at zero magnetic field. 
                  The inset shows the position of the maximum 
                  $k_{B}T_{\mathrm{max}}$ as a function of
                  $\hbar\omega_{c}$  with $g = 0$.}
    \label{fig:ResistivityT}
    \end{figure}

    Finally, we discuss the temperature dependence of the resistivity.
    Figure~\ref{fig:ResistivityT} shows $\rho(T,0)$ for several magnetic fields.
    The characteristic feature of resistivity in Fig.~\ref{fig:ResistivityT} is 
    the maximum in the temperature dependence which separates the high-temperature
    region ($k_{B}T \gg \ce$) from the low-temperature region ($k_{B}T \ll \ce$).
    In the high-temperature region, the resistivity is
    determined by the thermal activation of the carriers, which leads to the
    increase of the resistivity with lowering the temperature.
    In the low-temperature region, where the $n = 0$ level becomes isolated,
    the behavior of the resistivity is the opposite, since 
    increasing the temperature reduces the density of states 
    at $\mu = 0$ due to the temperature broadening. Therefore, resistivity 
    takes the maximum at $k_{B}T_{\mathrm{max}}\sim\ce$.
    The result of the numerical calculation of $T_{\mathrm{max}}$ on the
    $(\ce,k_{B}T)$-plane is presented in the inset in Fig.~\ref{fig:ResistivityT},
    which shows that at high $\ce$, $k_{B}T_{\mathrm{max}} \approx 0.21 (\ce - 0.5 \Gamma)$.
    In low magnetic fields, the maximum disappears when $\ce \approx 1.30\,\Gamma$, 
    which also gives zero for the right-hand side in Eq.~(\ref{d2sigma}).

    \subsection{Conductivity in SCBA}
    
    \begin{figure}
        \centering
        \begin{minipage}{.45\textwidth}
            \centerline{\includegraphics[scale=.75]{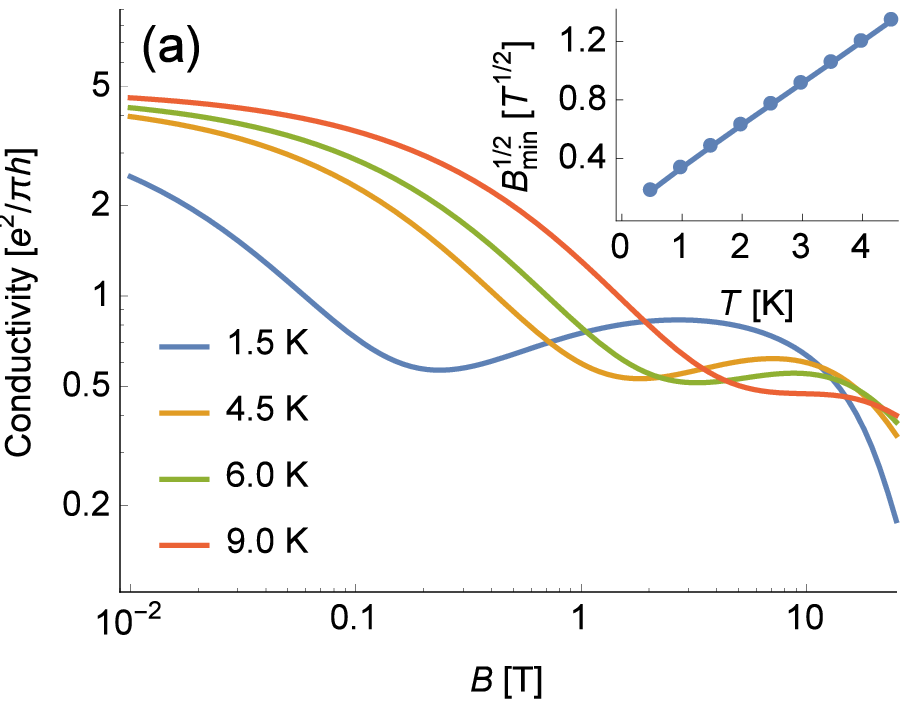}}
        \end{minipage}
        \begin{minipage}{.45\textwidth}
            \centerline{\includegraphics[scale=.75]{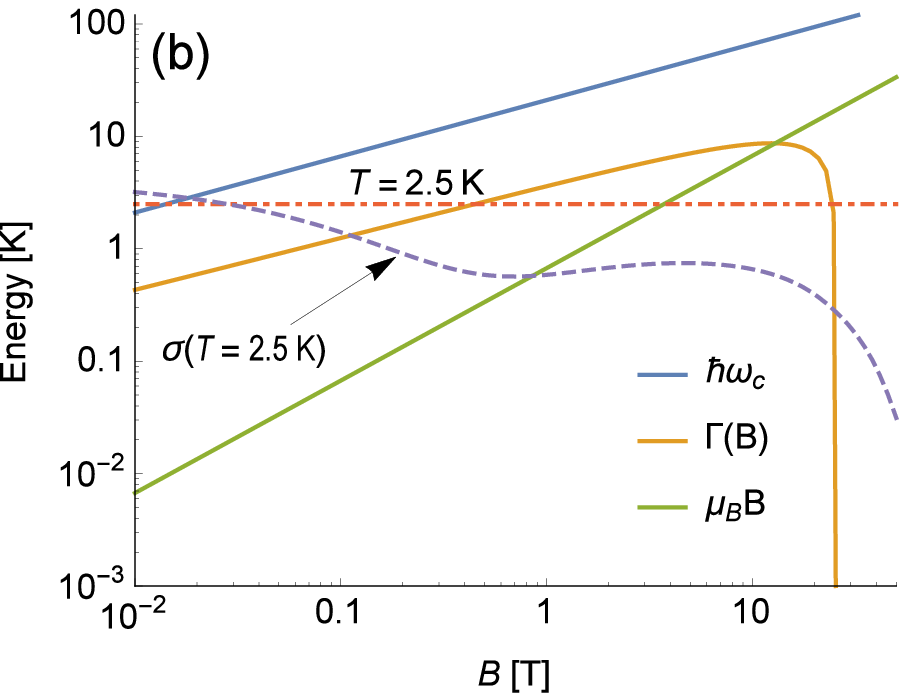}}
        \end{minipage}
        \caption{(Color online) 
                 (a) Magnetic-field dependence of the conductivity at $\mu = 0$
                 in SCBA for several temperatures with $\varkappa = 0.02$
                 and $E_{c} = 0.05$~eV. The inset shows $B_{\mathrm{min}}^{1/2}$ vs
                 temperature.
                 (b) Magnetic-field dependence of $\hbar \omega_{c}$, $\Gamma(B)$,
                 $\mu_{B} B$, and the  
                 conductivity in the unit of $e^2/\pi h$ at $T = 2.5$~K (dashed line)
                 in the same plot. The horizontal dotted-dashed line corresponds
                 to $T = 2.5$~K.}
    \label{fig:CondSCBA}
    \end{figure}
    
    In this section, we discuss how SCBA  
    modifies the results obtained in the previous section.
    The main feature in SCBA is that the level
    broadening $\Gamma(B)$ increases as $\sqrt{B}$.\cite{Shon1998} 
    As a result, the behavior of $\sigma$ changes when 
    $\Gamma(B) \sim k_{B}T$. 
    This leads to a minimum of the conductivity which is the central 
    result in this section.

    Figure~\ref{fig:CondSCBA}(a) shows the conductivity for $\mu = 0$
    and $\eta = 0$ with the Zeeman interaction
    as a function of the magnetic field in SCBA for several temperatures,
    where the self-energy was calculated numerically from Eq.~(\ref{eq:scba0}).
    In contrast to the simple two-step decrease in Fig.~\ref{fig3}(b),
    the conductivity in SCBA takes a  minimum in the moderate magnetic field 
    region. We find that this minimum is due to the crossover between 
    the ``temperature-dominating'' 
    [$\Gamma(B) \lesssim k_{B}T$] and ``impurity-dominating'' 
    [$k_{B}T \lesssim \Gamma(B)$] regimes. 
    In order to explain this, we show the energy scales of our problem, i.~e., 
    $\hbar\omega_{c}$, 
    $\Gamma(B)$ [Eq.~(\ref{GammaB})], $\mu_{B}B$, and $k_{B}T$ (dotted-dashed line) 
    in Fig.~\ref{fig:CondSCBA}(b).
    The dashed line is the conductivity for $T=2.5$~K 
    as a function of $B$. We can see that the minimum of $\sigma$
    occurs at $\Gamma(B) \sim k_{B}T$.
    In the low-magnetic field region, $\Gamma(B)$ is smaller than 
    $k_{B}T$, which corresponds to the temperature dominating region.
    In this case, $\sigma$ decreases according to Eq.~(\ref{eq54}).
    On the other hand, when $\Gamma(B) > k_{B}T$, the conductivity
    slightly increases. This is because $\sigma$  can be approximated
    from Eqs.~(\ref{eq:crossover}) and (\ref{d2sigma}) 
    as $\sigma = 1-\pi^{2}k_{B}^{2}T^{2}/3\Gamma^{2}(B)$ which approaches $1$
    with increasing $\Gamma(B)$. Then, in the high-magnetic field region,
    $\Gamma(B)$ starts to decrease, and, as a result, $\sigma$ decreases
    due to the Zeeman interaction.

    In order to confirm this interpretation, we calculate $B_{\mathrm{min}}$
    of various temperatures numerically. As shown in the inset of
    Fig.~\ref{fig:CondSCBA}(a),
    $B_{\mathrm{min}}$ scales as $T^{2}$
    in agreement with the relation $\Gamma(B) \sim \sqrt{B} \sim k_{B}T$.
    The minimum  of $\sigma$ disappears
    at high temperatures
    [e.~g., ``$T=9$~K'' curve in Fig.~\ref{fig:CondSCBA}(a)]. 
    This is because $k_{B}T > \Gamma(B)$ holds  in all range of magnetic
    fields. 
    
    In the high-magnetic-field region where the Zeeman term reduces $\sigma$,
    the $B$ dependence of $\sigma$ is different from that obtained in the previous section.
    At $T = 0$, according to Eq.~(\ref{eq:sigmaN}), we obtain
    \begin{equation}
        \sigma(0,0) = \left \lbrace
        \begin{array}{ll}        
        \frac{e^{2}}{\pi h} 
        \left (1 - \frac{B}{B_{c}} \right ),\quad & \mbox{for $B < B_{c}$},\\
        0, & \quad \mbox{otherwise},
        \end{array}
    \right.
    \end{equation}
    where $B_{c} =\varkappa e \hbar v_{F}^{2}/(g^{2}\mu_{B}^{2})$. 
    At finite temperatures, the conductivity in the region $B \gtrsim B_{c}$ 
    decreases exponentially with magnetic field,
    $\sigma(T) \sim \exp(-\mu_{B}B/k_{B}T)$.

    \begin{figure}
      \centerline{\includegraphics[scale=.75]{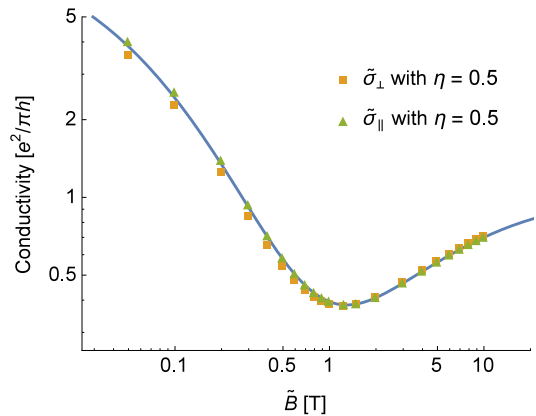}}
        \caption{(Color online) Magnetic-field dependence of the conductivity at $\mu = 0$
                 in SCBA for $T=4.5$~K, $\varkappa = 0.02$, and $E_{c} = 0.05$~eV
                 calculated from Eqs.~(\ref{eq:sigma03}) and (\ref{eq:scba0}) (solid line).
                 Closed squares (triangles) show $\tilde{\sigma}_{\perp}$ 
                 ($\tilde{\sigma}_{\parallel}$) with $\eta = 0.5$ calculated 
                 from Eqs.~(\ref{eq:sigmaT}) and (\ref{eq:s_per}) [Eq.~(\ref{eq:s_par})] using 
                 summation over 30 Landau levels. 
                 The horizontal axis is renormalized as $\tilde{B} = B (1-\eta^{2})^{-3/4}$.}
    \label{fig:tilting}
    \end{figure}

    Next, we study how our results for the magnetic-field dependence 
    of the conductivity are modified by the tilting of the Dirac cone.
    First of all, there are $\eta$-dependent prefactors of 
    $\sigma_{\perp}$ and $\sigma_{\parallel}$ [Eqs.~(\ref{eq:QNper}] and (\ref{eq:QNpar})].
    We will show that the magnetic-field dependencies of $\sigma_{\perp}$ and
    $\sigma_{\parallel}$ are not affected so much by $\eta$ except for 
    these prefactors.
    Figure~\ref{fig:tilting} shows the normalized conductivities 
    $\tilde\sigma_{\perp} = \lambda \sigma_{\perp}$ and
    $\tilde\sigma_{\parallel} = \lambda^{-1}\sigma_{\parallel}$ 
    as a function of $B$ for $T = 4.5$~K and
    $\varkappa = 0.02$ where there is no Zeeman term. Note that
    the magnetic field is also renormalized by $\eta$ as 
    $\ce \to \lambda^{-3/2} \ce$. Thus, the horizontal axis in Fig.~\ref{fig:tilting}
    is $\tilde{B} = B (1-\eta^{2})^{-3/4}$.
    Closed squares (triangles) in Fig.~\ref{fig:tilting} 
    show $\tilde{\sigma}_{\perp}$ ($\tilde{\sigma}_{\parallel}$) 
    calculated from Eqs.~(\ref{eq:sigmaT}) and (\ref{eq:s_per}) [Eq.~(\ref{eq:s_par})] using 
    numerical summation.
    The solid line is the same curve as in Fig.~\ref{fig:CondSCBA}(a) for $T = 4.5$~K
    which is shown 
    for comparison in the case of $\eta = 0$.
    Figure~\ref{fig:tilting} indicates that  $\eta$ has very small effects on the normalized 
    conductivities.
    Only in the weak magnetic-field region, 
    $\tilde{\sigma}_{\parallel}$ is slightly larger than $\tilde{\sigma}_{\perp}$.
    This difference can be understood in terms of Eqs.~(\ref{eq:QNper}) and (\ref{eq:QNpar}).
    In this case, there are Landau level mixings,  and
    the contribution from higher Landau levels for $\tilde{\sigma}_{\perp}$ is
    smaller by the factor $N\eta^{2}$ than that for $\tilde{\sigma}_{\parallel}$. 
    
    The physical reason of this small $\eta$ dependence will be as follows. 
    The basic elements of our explanation for the 
    magnetic-field dependence are the Landau level quantization and 
    the magnetic-field dependence of the level broadening.
    This explanation remains qualitatively
    the same even if the tilting is introduced. 
    Moreover, in the quantum limit, the conductivity is solely determined by $n = 0$ 
    Landau level. 
    In this case, $n = 0$ Landau level is 
    insensitive to $\eta$ except for the rescaling factors $\lambda$ and $\lambda^{-1}$.
    This will be the reason why the tilting has very small effect 
    on the magnetic-field dependence of the conductivity.
    Note that in the high-magnetic field region, where  
    $\sigma_{\perp}$ ($\sigma_{\parallel}$) reduces due to the Zeeman splitting
    of $n =0$ Landau level, the $\eta$ dependence comes only through the 
    prefactor $\lambda^{-1}$ ($\lambda$).

\begin{figure}
        \centerline{\includegraphics[scale=.75]{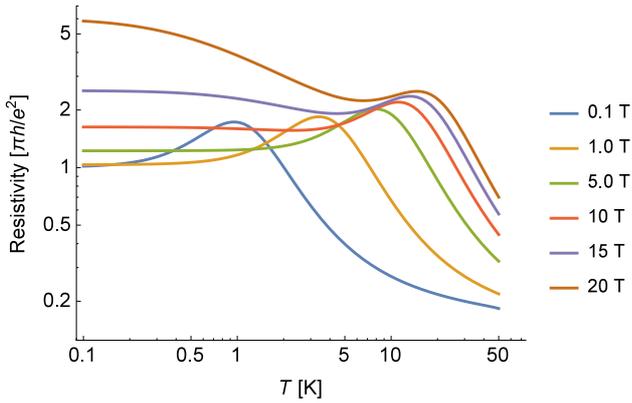}}
        \caption{(Color online) Resistivity at $\mu = 0$ as a function of temperature
                  for several magnetic fields calculated in SCBA with 
                  $\varkappa = 0.02$ and $E_{c} = 0.05$~eV. }
    \label{fig:RTscba}
    \end{figure}

    Finally, we briefly mention the temperature dependence of the resistivity in SCBA
    shown in Fig.~\ref{fig:RTscba} for several magnetic fields. 
    The explanation of this behavior remains qualitatively the same as in
    the previous section for the constant broadening approximation.
    The main difference of the present case from Fig.~\ref{fig:ResistivityT}
    is that the maximum in the temperature dependence of the resistivity survives 
    at small magnetic fields and becomes more pronounced as magnetic field decreases.  
    This behavior can be understood from Fig.~\ref{fig:gamma} which indicates that
    the $n = 0$ level becomes sharper when magnetic field goes to zero, since $\Gamma(B)$ 
    for $\varkappa = 0.02$ becomes exponentially small at $B = 0$.
    
\section{Summary}
\label{secV}

    In summary, we have made an analytical calculation of the longitudinal conductivity
    of two-dimensional massless fermions with a tilted Dirac cone in the framework of the 
    Kubo-Basting-St\v{r}eda linear-response formula in strong magnetic field.
    The main difference of the present case from the case of the isotropic Dirac cone is that the tilting
    introduces Landau level mixing which breaks usual selection 
    rules for the matrix elements of velocity operators  
    which leads to the existence of matrix elements  between the Landau levels with arbitrary $n$.

    First, we have analyzed the conductivity at zero temperature
    in the quantized magnetic field limit. 
    In this case, we found that the conductivities at $\mu = E_{N}$ are given by
    $\sigma_{xx} = \lambda^{-1}(\delta_{N0}+2N-N\eta^{2})$ and
    $\sigma_{yy} = \lambda(\delta_{N0}+2N)$. 
    The factors $\lambda$ and $\lambda^{-1}$ are purely geometrical,
    due to the ellipticity of the quasi-classical electron orbit.
    Apart from these factors, $\sigma_{yy}$ remains the same 
    as in the case without the tilting, while $\sigma_{xx}$ is reduced by $N\eta^{2}$.
    We have explained this result as follows.    
    The conductivity at $\mu = E_{N}$ is determined by 
    the quantum-mechanical average of $(\Delta r_{l})^{2}$ calculated 
    with respect to the $N$th Landau level wave function.
    If the tilting of the cone is in the $k_{y}$ direction in the energy-momentum
    space, the center of the $N$th eigenfunction is displaced by $x_{N}$ along the
    $x$ direction in the real space.
    We have proven that this displacement gives the reduction of $(\Delta x)^{2}$,
    while $(\Delta y)^{2}$ remains unchanged.
    We note that this fact is purely quantum effect and has no quasi-classical interpretation.
    
    Second, we studied how the zero-temperature conductivity at $\mu = 0$
    depends on magnetic field and impurity scattering.
    Using numerical calculations, we have demonstrated that
    $\sigma_{xx}$ at $\mu  = 0$ remains independent of magnetic field and short-ranged disorder,
    which is similar to the case without tilting,
    while $\sigma_{yy}$ shows a crossover from the
    value $\frac{e^{2}}{\pi h}\frac{\sin^{-1} \eta}{\eta}$ in zero magnetic field 
    to the value $\frac{e^{2}}{\pi h}\sqrt{1-\eta^{2}}$ in the strong-field limit.
    We propose that this fact can be used as an experimental evidence of the tilting of the Dirac cone.
    Indeed, there are two possible sources of the anisotropy of the 
    conductivity, namely, owing to the difference in the Fermi velocities
    $v_{x} \ne v_{y}$, and due to the tilting. In the former case,
    $\sigma_{yy}/\sigma_{xx} = v_{x}^{2}/v_{y}^{2} $ does not depend on
    magnetic field, while in the latter case $\sigma_{yy}/\sigma_{xx}$ changes
    from $\sqrt{1-\eta^{2}} \sin^{-1}\eta/\eta$ in zero magnetic field
    to $1-\eta^{2}$ in the high-field limit. For \bedt{} with $\eta \approx 0.8$,
    $\sigma_{yy}/\sigma_{xx}$ changes, respectively, from $0.69$ to $0.36$
    which can be detected experimentally. The magnetic-field dependent 
    anisotropy of the in-plane conductivity can be regarded as a characteristic feature
    of a tilted Dirac cone.

   Next, we have studied the conductivity at finite temperatures where
   we set $\eta = 0$ at the beginning.
   We have demonstrated that as a function of the magnetic field the conductivity at 
   the Dirac point undergoes a two-step decrease where the first 
   decrease is associated with the crossover to the quantum limit and
   the second decrease is due to the spin splitting of the $n =0$ level.
   For the effect of the impurities, the results are given for two models: 
   the constant broadening approximation and SCBA.
   The analysis based on the SCBA shows that the broadening of the Landau levels
   increases with the magnetic field as $\sqrt{B}$, which 
   leads to the minimum in the magnetoconductivity at $k_{B}T \sim \Gamma(B)
   \sim \sqrt{B_{\mathrm{min}}}$ 
   which separates the temperature dominated region $k_{B}T \gtrsim \Gamma(B)$ from the impurity
   dominated region $k_{B}T \lesssim \Gamma(B)$.
   We have demonstrated that this behavior remains qualitatively the same
   even in the presence of the tilting of the cone.

   The results of Sec~\ref{secIV} can be directly compared with recent experiments in 
   \bedt{} organic conductor. \cite{Monteverde2013,Ozawa2014}
   We note, that in contrast with  Ref.~\onlinecite{Monteverde2013},
   our analysis is able to explain the magnetic-field dependence of the conductivity in
   \bedt{} using only one type of Dirac carriers.
   Moreover, the minimum in the magnetoconductivity,
   which scales with temperature as $T^{2}$, can be regarded as a strong evidence
   in favor of the one-carrier scenario of the magnetotransport.
   
   In conclusion, we comment on the minor importance of the tilting of the Dirac cone
   for understanding the existent experiments in magnetic field in \bedt{}  
   near $\mu = 0$. We note that the key ingredients 
   of our explanation of the  magnetic-field dependence 
   of the conductivity, namely, the Landau quantization, the existence of the zero mode,
   and the magnetic-field dependence of the level broadening remain qualitatively 
   the same even in the presence of the tilting. That is the reason why our main results in 
   Sec.~\ref{secIV} such as two-step behavior of the conductivity in magnetic field and
   the minimum in the magneto-conductivity remain valid in the presence of the finite tilting.
   In order to detect the effect of the tilting on the in-plane transport,
   the authors propose to measure the magnetic-field dependence of
   $\sigma_{xx}/\sigma_{yy}$. This effect survives at finite temperatures 
   and can be used as a marker of a tilted cone.

\begin{acknowledgments}
    The authors are thankful to N.~Tajima for fruitful discussions.
    One of the authors (Y.S.) also thanks M.~Monteverde 
    and M.~O.~Goerbig for useful comments.
    This work was supported by Grants-in-Aid for Scientific Research 
    (A) (No.~24244053) and 
    (C) (No.~26400355)
    from the Ministry of Education, 
    Culture, Sports, Science and Technology of Japan. 
    One of the authors (I.~P.) acknowledges the Russian Foundation for 
    Basic Research, 
    Grant No.~14-02-92104 and 
    the Government of the Russian Federation, Program No.~02.A03.21.0006.
\end{acknowledgments}

\appendix

\section{Matrix elements of velocity operators}
\label{Appendix_A}

    The matrix elements of the velocity operators in Eqs.~(\ref{eq:vx}) and (\ref{eq:vy})
    for the Landau levels in Eq.~(\ref{eq:eigenfunc}) can be expressed in the following form:
    \begin{eqnarray}
        \braket{m}{\bar{v}_{x}}{n} &=& i\lambda C_{nm} 
        \left [\sgn(m) F_{nm}^{(01)} - \sgn(n) F_{nm}^{(10)} \right ], 
    \label{app:vx}
    \\
    \braket{m}{\bar{v}_{y}}{n} &=& C_{nm}\left [
    -\eta F_{nm}^{(00)}+\sgn(n) F_{nm}^{(10)} + \sgn(m) F_{nm}^{(01)} \right . \nonumber   \\
    &-& \left . \eta \sgn(m)\sgn(n) F_{nm}^{(11)} \right ],
    \label{app:vy}
    \end{eqnarray}
    where $C_{nm} = (2-\delta_{n0})^{-1/2}(2-\delta_{m0})^{-1/2}$,
    \begin{equation}
        F_{n,m}^{(ij)} = \int_{-\infty}^{\infty} dx
        h_{|n|-i}(x-x_{n}) h_{|m|-j}(x-x_{m}),
    \end{equation}
    $i,j=0,1$, and $|n|\ge i$, $|m| \ge j$. 
    The coefficients $F_{nm}^{(ij)}$ can be calculated 
    using the formula \cite{Gradshteyn2007}
    \begin{multline}
        \int_{-\infty}^{\infty} e^{-x^{2}} H_{m}(x+y) H_{n}(x+z) dx \\
        = 2^{n} \pi^{1/2} m! z^{n-m} L_{m}^{n-m}(-2yz), \quad (m \le n),
    \end{multline}
    which gives
    \begin{eqnarray}
        F_{nm}^{(00)} &=& N_{n}^{m} L_{|m|}^{|n|-|m|}(\eta^{2} \Delta_{n,m}^{2} ),	
    \label{app:f00} 
    \\
        F_{nm}^{(10)} &=& N_{n}^{m} \frac{\sqrt{|n|}}{\eta \Delta_{n,m}} 
	    L_{|m|}^{|n|-|m|-1}(\eta^{2} \Delta_{n,m}^{2} ),	
	\\
        F_{nm}^{(01)} &=& N_{n}^{m} \frac{\eta \Delta_{n,m}}{\sqrt{|m|}}
	    L_{|m|-1}^{|n|-|m|+1}(\eta^{2} \Delta_{n,m}^{2} ),	
	\\
        F_{nm}^{(11)} &=& N_{n}^{m} \sqrt{\frac{|n|}{|m|}}
        L_{|m|-1}^{|n|-|m|}(\eta^{2} \Delta_{n,m}^{2} ),  
    \label{app:f11}
    \end{eqnarray}
    where
    \begin{equation}
        N_{n}^{m} = \sqrt{\frac{|m|!}{|n|!}} \left ( \eta \Delta_{n,m} \right )^{|n|-|m|}
        \exp\left ( -\frac{\eta^{2}}{2}  \Delta_{n,m}^{2} \right ),
    \end{equation} 
    and we used the property of Laguerre polynomials with integer $n$ and $m$:
    \begin{equation}
        \frac{(-x)^{m}}{m!} L_{n}^{m-n}(x) = \frac{(-x)^{n}}{n!}L_{m}^{n-m}(x).
    \end{equation}
    Substituting Eqs.~(\ref{app:f00})--(\ref{app:f11}) in the matrix elements~(\ref{app:vx}) and
    (\ref{app:vy}) and using the recurrence relations for Laguerre polynomials
    \begin{equation}
    L_{|m|}^{|n|-|m|-1}(x) = L_{|m|}^{|n|-|m|}(x)-L_{|m|-1}^{|n|-|m|}(x),
    \end{equation}
    \begin{equation}
    x L_{|m|-1}^{|n|-|m|+1}(x) = |n| L_{|m|-1}^{|n|-|m|}(x)-|m|L_{|m|}^{|n|-|m|}(x),
    \end{equation}
    we obtain the expressions (\ref{eq:vx}) and (\ref{eq:vy}).

\onecolumngrid


\section{Conductivity in the $\eta \to 0$ limit}
\label{Appendix_B}
   
    If $\eta = 0$, the matrix elements in Eqs.~(\ref{eq:vx}) and (\ref{eq:vy}) 
    become
    \begin{eqnarray}
        \braket{m}{\bar{v}_{x}}{n} &=& 
        -i C_{nm} \left [\sgn(n) \delta_{|n||m|+1} - \sgn(m) \delta_{|m||n|+1}\right ], 
        \\
        \braket{m}{\bar{v}_{y}}{n} &=&
        \phantom{-i} C_{nm} \left [ \sgn(n) \delta_{|n||m|+1} + \sgn(m) \delta_{|m||n|+1} \right ],
    \end{eqnarray}
    where $C_{nm} = (2-\delta_{n0})^{-1/2}(2-\delta_{m0})^{-1/2}$.
    In this case, Eqs.~(\ref{eq:s_per}) and (\ref{eq:s_par}) for the conductivities
    reduce to Eq.~(\ref{eq:sigma02}). 
    In order to make a summation over $n$ in
    Eq.~(\ref{eq:sigma02}), we make a summation over $\alpha$ and 
    $\alpha'$ with the help of the identity
    \begin{equation}
        \frac{\gamma}{(x - \alpha \sqrt{n})^{2}+\gamma^{2}} = \frac{1}{2i} \left ( 
        \frac{1}{x - \alpha \sqrt{n}-i\gamma}-\frac{1}{x - \alpha \sqrt{n}+i\gamma} \right ),
    \end{equation}
    which yields
    \begin{equation}
        \sigma_{0}(x) = \frac{1}{4} \sum_{n=0}^{\infty}
        \left ( \frac{\gamma_{+}}{n+1+\gamma_{+}^{2}} + \frac{\gamma_{-}}{n+1+\gamma_{-}^{2}} \right )
        \left ( \frac{\gamma_{+}}{n+\gamma_{+}^{2}} + \frac{\gamma_{-}}{n+\gamma_{-}^{2}} \right )
    \end{equation}
    where we introduced a shorthand notation $\gamma_{\pm} = (\gamma \pm i x )/\ce$.
    Now the summation over $n$ can be easily done with the help of the formula
    \begin{equation}
        \sum_{n=0}^{\infty} \frac{1}{(n+a)(n+b)} = \frac{1}{b-a} \left[ \psi(b) - \psi(a) \right],
    \end{equation}
    where $\psi$ denotes digamma function. 
    The final result is
    \begin{equation}
    \sigma_{0}(x) = \frac{1}{2} \left [ 
    1 + \frac{\gamma_{+}^{2}+\gamma_{-}^{2}+(\gamma_{+}^{2}-\gamma_{-}^{2})^{2}}
    {2\gamma_{+}\gamma_{-}\left ( 1 - (\gamma_{+}^{2}-\gamma_{-}^{2})^{2} \right ) }-
    \frac{\gamma_{+}\gamma_{-}\left ( \gamma_{+}^{2} - \gamma_{-}^{2} \right )}
    {1 - (\gamma_{+}^{2}-\gamma_{-}^{2})^{2}}
    \left (\psi(\gamma_{+}^{2})-\psi(\gamma_{-}^{2}) \right ) \right ]
    \end{equation}
    which gives Eq.~(\ref{eq:sigma03}).

    In the limit $\eta \Delta_{n,m} \to 0 $ we can expand
    Eqs.~(\ref{eq:vx}) and (\ref{eq:vy}) in powers of $\eta$.
    Up to the order of $\eta^{2}$, $v_{x}$ and $v_{y}$ have matrix elements up to
    $|m| = |n| \pm 3$. 
    The expansion has the form
    \begin{equation}
        \braket{m}{\bar{v}_{x}}{n} =
        i \lambda C_{nm} \sum_{j=1}^{3} 
        \left [ a_{n}^{(j)} \delta_{|m|=|n|+j} -  
         a_{m}^{(j)} \delta_{|n|=|m|+j} \right ] + O(\eta^{3}),
    \label{app:vx_expand}
    \end{equation}
    where
    \begin{eqnarray}
        a_{n}^{(1)} &=& \sgn(m) \left \lbrace 1 - \frac{1}{2}\left ( \sqrt{|n|+1} - \sgn(nm) \sqrt{|n|}\right )
        \left [ \left (|n|+1\right )^{3/2} - \sgn(nm) |n|^{3/2} \right ] \eta^{2} \right \rbrace, 
        \\
        a_{n}^{(2)} &=& \sqrt{|n|+1} \left ( \sqrt{|n|+2} -\sgn(nm) \sqrt{|n|} \right ) \eta, 
        \\
        a_{n}^{(3)} &=& \frac{1}{2}\sgn(m) \sqrt{(|n|+1)(|n|+2)} 
        \left (\sqrt{|n|+3} -\sgn(nm) \sqrt{|n|}\right )^{2} \eta^{2},
    \end{eqnarray}
    and
    \begin{equation}
        \braket{m}{\bar{v}_{y}}{n} =
        \lambda^{2} C_{nm} \sum_{j=0}^{3} 
        \left [ b_{n}^{(j)} \delta_{|m|=|n|+j} +  
                b_{m}^{(j)} \delta_{|n|=|m|+j} \right ] + O(\eta^{3}),
    \label{app:vy_expand}
    \end{equation}
    where
    \begin{eqnarray}
    b_{n}^{(0)} &=& -2 n \eta,
    \\
    b_{n}^{(1)} &=& \sgn(m)\left \lbrace 1 - \frac{1}{2}\left ( \sqrt{|n|+1} - \sgn(nm) \sqrt{|n|}\right )^{3}
    \left [ \left (|n|+1\right )^{3/2} + \sgn(nm) |n|^{3/2} \right ] \eta^{2} \right \rbrace, 
    \\
    b_{n}^{(2)}&=& a_{n}^{(2)},\quad b_{n}^{(3)} = a_{n}^{(3)}.
\end{eqnarray}

\section{Calculation of $\langle (\Delta r_{l})^{2} \rangle_{N}$}
\label{Appendix_C}

In the Landau gauge, which is used in this paper, the $x$ operator is bounded,
and calculation of the following matrix elements is straightforward.
\begin{multline}
\braket{\Psi_{kN}}{\left (x - x_{N}\right )^{p}}{\Psi_{k'N}} = 
\frac{\delta_{kk'}\ell^{2}}{2-\delta_{N0}}
\left\lbrace \int_{-\infty}^{\infty} dx h_{|N|}^{2}(x)x^{p}  \right.\\\left.
 -2 \sgn(n)\eta \int_{-\infty}^{\infty} dx 
h_{|N|}(x)h_{|N|-1}(x)x^{p}
 + (1-\delta_{N0})\int_{-\infty}^{\infty} dx h_{|N|-1}^{2}(x)x^{p}  \right\rbrace
\end{multline}
where $p = 1,2$, and the integrals are calculated as
\begin{equation}
\int_{-\infty}^{\infty} dx h_{|N|}(x) x^{p} = \frac{\delta_{p2}}{2\lambda}
\left ( 2|N| + 1 \right ), \qquad \int_{-\infty}^{\infty} dx 
h_{|N|}(x)h_{|N|-1}(x)x^{p} = \delta_{p1} \sqrt{\frac{|N|}{2\lambda}},
\end{equation}
which justifies Eqs.~(\ref{eq:x2}) and (\ref{eq:x}). However, the $y$ operator is unbounded, 
which usually means that the $y$ integral is ill defined and 
depends on the choice of the limits. We choose the following regularization. We add 
$\exp\left (-\frac{\epsilon^{2}}{4} y^{2} \right )$ where $\epsilon \to 0+$, and, after
that, expand the integration limits to infinity:
\begin{eqnarray}
\int_{-\infty}^{\infty} dy \exp\left [i(k-k')y - \frac{\epsilon^{2}}{4}y^{2}\right ] &=&
2\pi \frac{\exp\left [ -\frac{(k-k')^{2}}{\epsilon^{2}} \right ]}{\sqrt{\pi}\epsilon} \to
2 \pi \delta(k-k'), \\
\int_{-\infty}^{\infty} dy \exp\left [i(k-k')y - \frac{\epsilon^{2}}{4}y^{2}\right ] y &=&
-2\pi i \frac{d}{dk} \frac{\exp\left [ -\frac{(k-k')^{2}}{\epsilon^{2}} \right ]}{\sqrt{\pi}\epsilon}
\to -2\pi i \delta'(k-k'),\\
\int_{-\infty}^{\infty} dy \exp\left [i(k-k')y - \frac{\epsilon^{2}}{4}y^{2}\right ] y^{2} &=&
-2\pi \frac{d^{2}}{dk^{2}} \frac{\exp\left [ -\frac{(k-k')^{2}}{\epsilon^{2}} \right ]}
{\sqrt{\pi}\epsilon} \to -2\pi \delta''(k-k').
\end{eqnarray}
The  matrix elements of the $y$-operator correspond to the integrals of total derivatives of 
$h_{|N|}^{2}(x)$ and $h_{|N|}(x)h_{|N|-1}(x)$, which give zero. The matrix elements of $y^{2}$
can be expressed as
\begin{multline}
\braket{\Psi_{kN}}{y^{2}}{\Psi_{k'N}} = \frac{\delta_{kk'}\ell^{2}}{2-\delta_{N0}}
\left \lbrace 
\int_{-\infty}^{\infty} dx \left ( h'_{|N|}(x) \right )^{2} - \right .\\
\left . 2 \sgn(n) \eta \int_{-\infty}^{\infty} dx h'_{|N|}(x) h'_{|N|-1}(x)
+ (1 - \delta_{N0}) \int_{-\infty}^{\infty} dx \left ( h'_{|N|-1}(x) \right )^{2}
\right \rbrace,
\end{multline}
where
\begin{equation}
\int_{-\infty}^{\infty} dx \left ( h'_{|N|}(x) \right )^{2} = \frac{\lambda}{2}(2|N|+1),
\quad \mbox{and} \quad
\int_{-\infty}^{\infty} dx h'_{|N|}(x) h'_{|N|-1}(x) = 0,
\end{equation}
which yields Eq.~(\ref{eq:y2}).

\twocolumngrid

\bibliography{References}{}

\end{document}